\documentclass[aps,epsfig]{revtex4}
\usepackage{graphicx}
\begin{document}
\def\be{\begin{equation}}
\def\ee{\end{equation}}
\def\bfi{\begin{figure}}
\def\efi{\end{figure}}
\def\bea{\begin{eqnarray}}
\def\eea{\end{eqnarray}}

\title{Influence of thermal fluctuations on the geometry of the interfaces  
of the quenched Ising model}

 \author{Federico Corberi}
\affiliation {Dipartimento di Matematica ed Informatica, Universit\`a di Salerno, 
via Ponte don Melillo, 84084 Fisciano (SA), Italy.}
\author{Eugenio Lippiello}
\affiliation {Dipartimento di Scienze Fisiche, Universit\'a di Napoli
``Federico II'', 80125 Napoli, Italy.}
\author{Marco Zannetti}
\affiliation {Dipartimento di Matematica ed Informatica, Universit\`a di Salerno, 
via Ponte don Melillo, 84084 Fisciano (SA), Italy. }

\begin{abstract}
We study the role of the quench temperature $T_f$ in the phase-ordering 
kinetics of the Ising model with single spin flip in $d=2,3$.
Equilibrium interfaces are flat at $T_f=0$, whereas at $T_f>0$ they are
curved and rough (above the roughening temperature in $d=3$). 
We show, by means of scaling arguments and numerical
simulations, that this geometrical difference is important for the phase-ordering kinetics as well.
In particular, while the growth exponent $z=2$ of the
size of domains $L(t)\sim t^{1/z}$ is unaffected by $T_f$, 
other  exponents related to the interface geometry take different values
at $T_f=0$ or $T_f>0$. For $T_f>0$ a crossover phenomenon is observed
from an early stage where interfaces are still flat and the system behaves as at $T_f=0$, to the asymptotic regime with curved interfaces characteristic
of $T_f>0$. Furthermore, it is shown that the roughening length, although
sub-dominant with respect to $L(t)$, produces appreciable correction to scaling
up to very long times in $d=2$.

\end{abstract}

\maketitle


\section{Introduction} \label{intro}

When a binary system in suddenly quenched from above the 
critical temperature $T_c$ to a temperature $T_f<T_c$,
phase-ordering occurs with 
formation and growth of domains. After a certain time
$t_{sc}$ dynamical scaling~\cite{Bray94} sets in,
characterized by the typical size of ordered regions
growing algebraically in time, $L(t)\sim t^{1/z}$. 
When domains are large, their bulk 
is in quasi-equilibrium in one of the two 
broken symmetry phases which are  
characterized by finite correlation length $\xi (T_f)$
and relaxation time $t_{eq}(T_f)\sim \xi ^z(T_f)$,
while the motion of the boundaries keeps the system
globally out of equilibrium. 
At a given time $s$, therefore,  
non-equilibrium effects can be detected
by looking over distances larger than $L(s)$, because in this case
one or more interfaces will be observed. On the other hand
a local observation performed from time $t=s$ onwards can reveal 
non-equilibrium features only for time separations $t-s>s$, because on these
timescales at least one interface has typically passed across the observation region. 
In the other regime, instead, for space separations $r \ll L(s)$ or time separations $t-s
\ll s$, the equilibrium properties of the interior of
domains are probed. 
This character of the dynamics induces an additive structure for 
pair correlation functions between local observables.   
Using the terminology of spin systems, and considering, for simplicity,
the spin-spin correlation function 
$G(r,t,s)=\langle \sigma_i(t)\sigma _j(s)\rangle -
\langle \sigma_i(t)\rangle \langle \sigma _j(s)\rangle $,
where $\sigma _i(t)$ is the value of the spin on site $i$ at time $t$ and
$r$ is the distance between sites $i,j$, one has
\be
G(r,t,s)=G_{st}(r,t-s)+G_{ag}(r,t,s).
\label{split}
\ee
The stationary term $G_{st}$ 
describes equilibrium fluctuations inside domains, and decays 
to zero for distances $r \gg \xi (T_f)$ and/or
time separations $t-s \gg t_{eq}(T_f)$.
$G_{ag}$, which contains the out of equilibrium
information, is the correlation function of
interest in the theory of phase ordering, and obeys the scaling 
form~\cite{Furukawa}
\be
G_{ag}(r,t,s)=\widehat{G}(r/L(s),t/s).
\label{1.132}
\ee
Furthermore, the scaling function for large time separation~\cite{Bray94}
is of the form
\be
\widehat{G}(r/L(s),t/s)\sim(t/s)^{-\lambda/z}h[r/L(s)]
\label{1.1321}
\ee
where $\lambda$ is the
Fisher-Huse exponent~\cite{Fisher88}, and, in system with
sharp interfaces, like the Ising model, the function $h(x)$ obeys
the Porod law
\be
1-h(x)\sim x
\label{porod}
\ee
for $x\lesssim 1$.
In general, both the terms of the splitting~(\ref{split})
display a {\it universal } character. For the stationary part, 
this is well known from equilibrium statistical mechanics,
where the renormalization group allows the classification of different
systems into universality classes on the basis of few relevant
parameters~\cite{hohenberg}.
A similar property is believed to hold also for the aging term. 
Universal indices, such as
the exponents $\lambda $, $z$, or other appearing in different 
quantities, should depend only on a small set of parameters
among which the space dimension $d$, the number of order parameter
components and the presence of conservation
laws in the dynamics. The theoretical foundations of this idea,
however, are not as robust as for its equilibrium counterpart
This is due to the non-perturbative character of the dynamical
problem. Actually, while in equilibrium an upper critical dimension
$d_U$ exists above which the renormalization group (RG) fixed point is
Gaussian, allowing the $\epsilon $-expansion for $d<d_U$, 
there is not an upper critical dimension for the dynamical 
process following a quench below $T_c$~\cite{mazenkorg}. Although an approach based fully on the RG is not available, 
complementing RG techniques with
a physically motivated ansatz, it has been shown~\cite{Bray89} 
that, for a system of continuous spins described by a time-dependent
Ginzburg-Landau (TDGL) equation
there exists an attractive 
strong coupling fixed point at $T=0$
governing the large scale properties of quenches to every
$T_f<T_c$. 
This result supports the idea of a universal
character of the aging term in Eq.~(\ref{split}), allows for a definition of non-equilibrium
universality classes and shows that universal quantities, such as exponents,
are the same in the whole low temperature phase.        
Restricting from now on to scalar systems
with short range interactions and
without dynamical conservation of the order parameter~\cite{Bray94},
these quantities should depend only on space dimension~\cite{sicilia}.
This would agree with the physical idea that
$T_f$ only determines the size $\xi (T_f)$ of the thermal
island of reversed spins inside the domains,
described by the stationary term in Eq.~(\ref{split}), leaving unchanged large-scale
long-time properties of the interface motion, contained
in the aging part. Basically, this indicates
a unique mechanism governing the non-equilibrium behavior
of interfaces. 
Restricting our attention to the exponent $z$, the Lifschitz-Cahn-Allen
theory~\cite{Allen79} confirms this idea, since it
gives $z=2$ for every $T_f<T_c$.
At the basis of this result is the so called
{\it curvature driven} mechanism: the existence of a surface
tension implies a force per unit of domains boundary area proportional
to the mean curvature which, in turn, is proportional to the inverse of $L(t)$.
For purely relaxational dynamics, this readily gives 
$z=2$, independent on $T_f$ and on dimensionality.    

These results are all based on continuous models
where the usual tools of differential analysis can be used
and the curvature is a well defined object. This approach is justified 
also for lattice models, 
such as the nearest neighbor Ising model, at relatively
high $T_f$, where temperature fluctuations produce  
soft interfaces, which at a coarse-grained level have a continuous character,
and can be well described in terms of partial differential equations.
When the temperature is lowered, however, these interfaces 
become faceted. This means that, although the growing structure
has still a bicontinuous interconnected morphology, 
interfaces are flat up to scales of order $L(t)$.
This implies that their description 
in the continuum may be inappropriate.
Then, while continuum theories predict temperature to be
an irrelevant parameter, with $T_f$-independent exponents
and a common kinetic mechanism for all quenches to
$T_f<T_c$, lattice models could in principle
behave differently,
in particular at $T_f=0$. This would imply that
temperature fluctuation do play a significant role
in the way interfaces evolve, determining, besides
the properties of the stationary term in Eq.~(\ref{split}), 
also those of the aging contribution.
This issue is not yet clarified;
let us mention, for example, that while for quenches to 
$T_f=0$ in $d=3$ the exponent $z=2$
has been observed~\cite{Brown02} in numerical simulations of 
the TDGL equation, for the Ising model one measures~\cite{Amar89}
an higher value whose origin is not yet clear. 
  
In this Paper we consider the role of $T_f$ in the phase-ordering
kinetics of the nearest neighbor Ising model.
For quenches to $T_f=0$ we will argue in Sec.~\ref{scal2d} that the basic mechanism
for the growth of $L(t)$ can be 
properly seen as a progressive elimination of small
domains with a faceted geometry, and that the zero temperature
constraint, namely the unrealizability of 
activated moves, plays a crucial role. Elaborating on this
we develop a scaling argument which allows us to determine 
analytically the behavior of several quantities. The results
of this approach are compared in Sec.~\ref{numerical} with the outcome of numerical
simulations, providing a general agreement. In particular, for the total interface density $\rho (t)$,
which is related to the domains size by
$\rho (t)^{-1}\propto L(t)$, we find
a power law behavior with $z=2$ in every dimension, as
at finite temperature. 
The different character of
the dynamics at $T_f=0$ is enlightened in Sec.˜\ref{densities}
by considering the densities $\rho _n (t)\propto L(t)^{-\beta_n}$ of spins $\sigma _i$
with a given {\it degree of alignment} $n$, this quantity being 
the difference between the
number of aligned and that of anti-aligned neighbors of $\sigma _i$. 
These quantities provide information
on the geometry of the interface and are shown
to behave differently
for quenches to finite $T_f$ or to $T_f=0$. 
While for shallow quenches, when the
curvature driven mechanism is at work, 
one has $\beta_n =1$ for every $n$,
for quenches to $T_f=0$ one finds $n$-dependent (and $d$-dependent) values
of $\beta_n $.  
For deep quenches with $T_f>0$, a crossover
is numerically observed (Sec.˜\ref{tneq0}) between an early stage (that can be rather
long for small $T_f$) where the same behavior of quenches to 
$T_f=0$ is observed, to the late regime dominated
by the usual curvature mechanism.

Finally, we discuss the effects of temperature fluctuations on the
characteristic time $t_{sc}$ of the onset of scaling. 
Our numerical simulations (Sec.~\ref{numerical}) show that the behavior of $t_{sc}$ is very different
in $d=2$ and in $d=3$. In $d=2$, 
$t_{sc}$ is relatively small in quenches to $T_f=0$
and grows monotonously raising $T_f$. 
For $T_f>0$ and $t<t_{sc}$ one observes an approximate power law behavior
with $L(t)\propto t^{1/z_{eff}(t)}$, with an effective exponent 
$z_{eff}(t)>2$, slowly converging to the asymptotic value.
This explains why values of $1/z\simeq 0.47-0.48$ are often 
reported in the literature~\cite{Manoy00}. We interpret the increase of $t_{sc}$
as due to the presence of the roughening length
$U(t,T_f)\propto t^{1/4}$ competing with $L(t)$ in the early regime.
This interpretation is shown to agree with the results of
numerical simulations. 
Moreover, we show how the effect of roughness can also be detected numerically in the behavior of $h(x)$
(Eqs.~(\ref{1.1321},\ref{porod})).
Actually, over distances $r<U(t,T_f)$ interfaces are not sharp, so that
the Porod law~(\ref{porod}) is not obeyed for $x<x_R(t)\simeq U(t,T_f)/L(t)\sim t^{-1/4}$.
For $d=3$, instead, we find the opposite situation, $t_{sc}$ is very large
at $T_f=0$, while it is small for shallow quenches. In $d=3$,
$U(t,T_f)$ grows at most logarithmically and hence is dominated
by $L(t)$ very soon causing no delays to scaling. Therefore, the mechanism
leading to the increase of $t_{sc}$ when raising $T_f$ in $d=2$ is not
present here and, in shallow
quenches, $t_{sc}$ is relatively small.
Instead, when $T_f=0$ numerical simulations show a very long lasting transient.
This is probably due to the constrained character of the kinetics where
activated moves are forbidden. 
The very large value of $t_{sc}$ explains the anomalous values of $1/z\simeq 0.33-0.37$
sometimes reported in the literature~\cite{Amar89} for quenches to $T_f=0$. 
However, our simulations
show unambiguously that $1/z_{eff}(t)$ is a growing function
of $t$ and, although at the longest simulated times it is still
$1/z_{eff}(t)\simeq 0.43$, its behavior is consistent
with an asymptotic value $z=2$.  

This paper is organized as follows: In Sec.\ref{scal2d} we define the
Ising model and develop scaling arguments to determine the behavior of 
$\rho (t)$ and $\rho _n(t)$ in quenches to $T_f=0$.
In Sec.~\ref{numerical} we discuss the data from simulations of quenches
to $T_f=0$ (Sec.~\ref{numteq0}) and to $T_f>0$ (Sec.~\ref{tneq0}), showing the agreement with
the results obtained in Sec.~\ref{scal2d}. Sec.~\ref{concl}
contains the conclusions.

\section{Scaling arguments for quenches to $T_f=0$.}\label{scal2d}

In the following we will consider the nearest neighbor Ising model
described by the Hamiltonian
\be
H([\sigma ])=-J\sum _ {\langle i,j \rangle } \sigma _i \sigma _j 
\ee
where $\sigma _i=\pm 1$ are the spin variables, $\langle i,j \rangle$ 
are two nearest sites on a $d$-dimensional lattice and $[\sigma ]$ is
the configuration of all the spins.
A purely relaxational dynamics without conservation of the order parameter
can be defined by introducing single spin flip transition rates
$w(\sigma _i\to -\sigma _i)$ obeying detailed balance.
These quantities depend on $T_f$ and on the local energy  
$E_i=-Jn_i$, $n_i$ being the {\it degree of alignment}, namely the difference between the
number of the neighboring spins  aligned with $\sigma _i$ 
and that of the anti-aligned ones. Letting $J=1$, transition rates
are functions of $n_i$ and $T_f$, namely 
$w(\sigma _i\to -\sigma _i)=W(n_i[\sigma ],T_f)$.
For $T_f=0$ one has $W(n_i[\sigma],T_f)=0$ whenever $n_i>0$.

In the remaining of this Section  we will develop a scaling argument
to determine the behavior of several quantities, among which $L(t)$,
in quenches to $T_f=0$.

\begin{figure}
\vspace{2cm}
    \centering
   \rotatebox{0}{\resizebox{.45\textwidth}{!}{\includegraphics{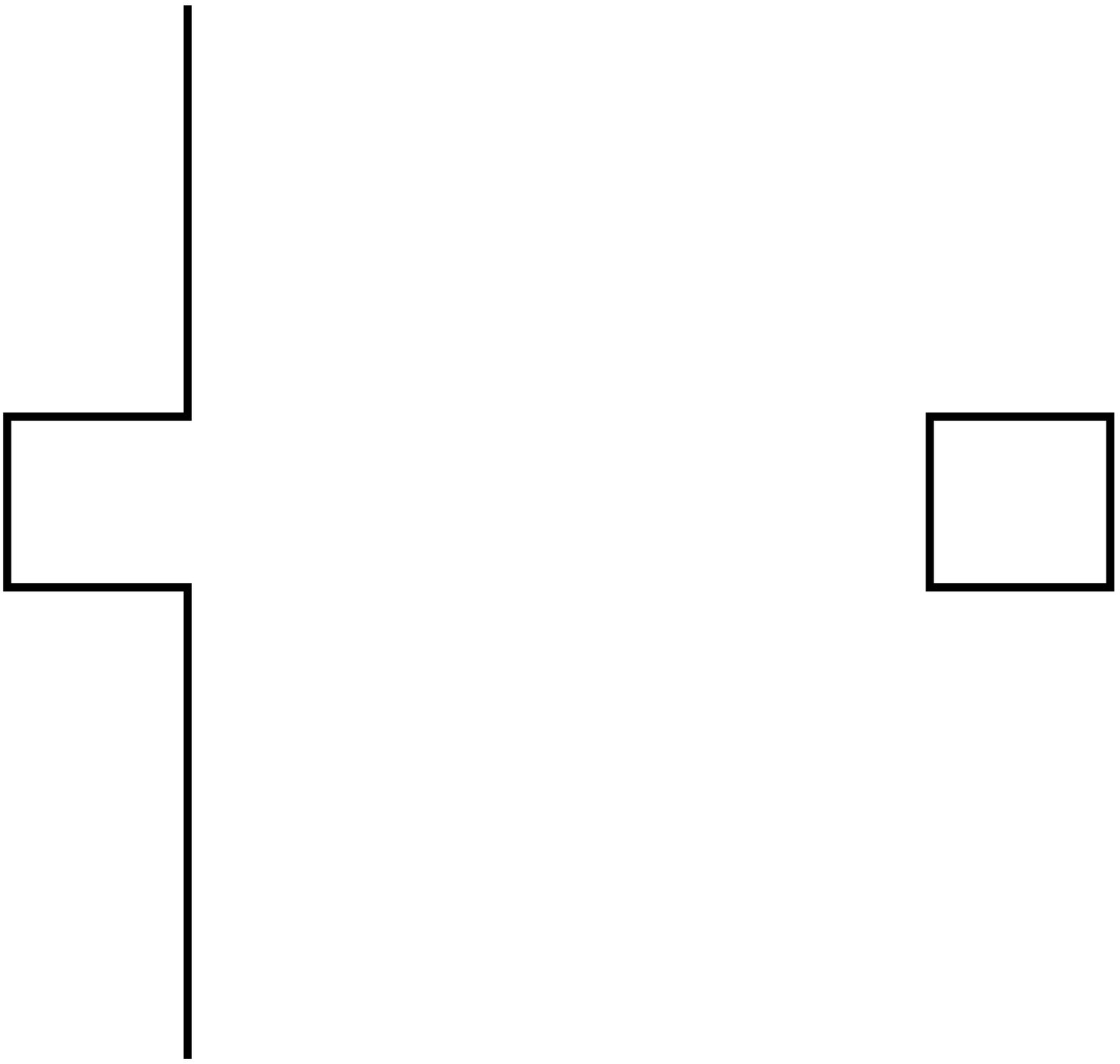}}}

   \vspace{2cm}
   \rotatebox{0}{\resizebox{.45\textwidth}{!}{\includegraphics{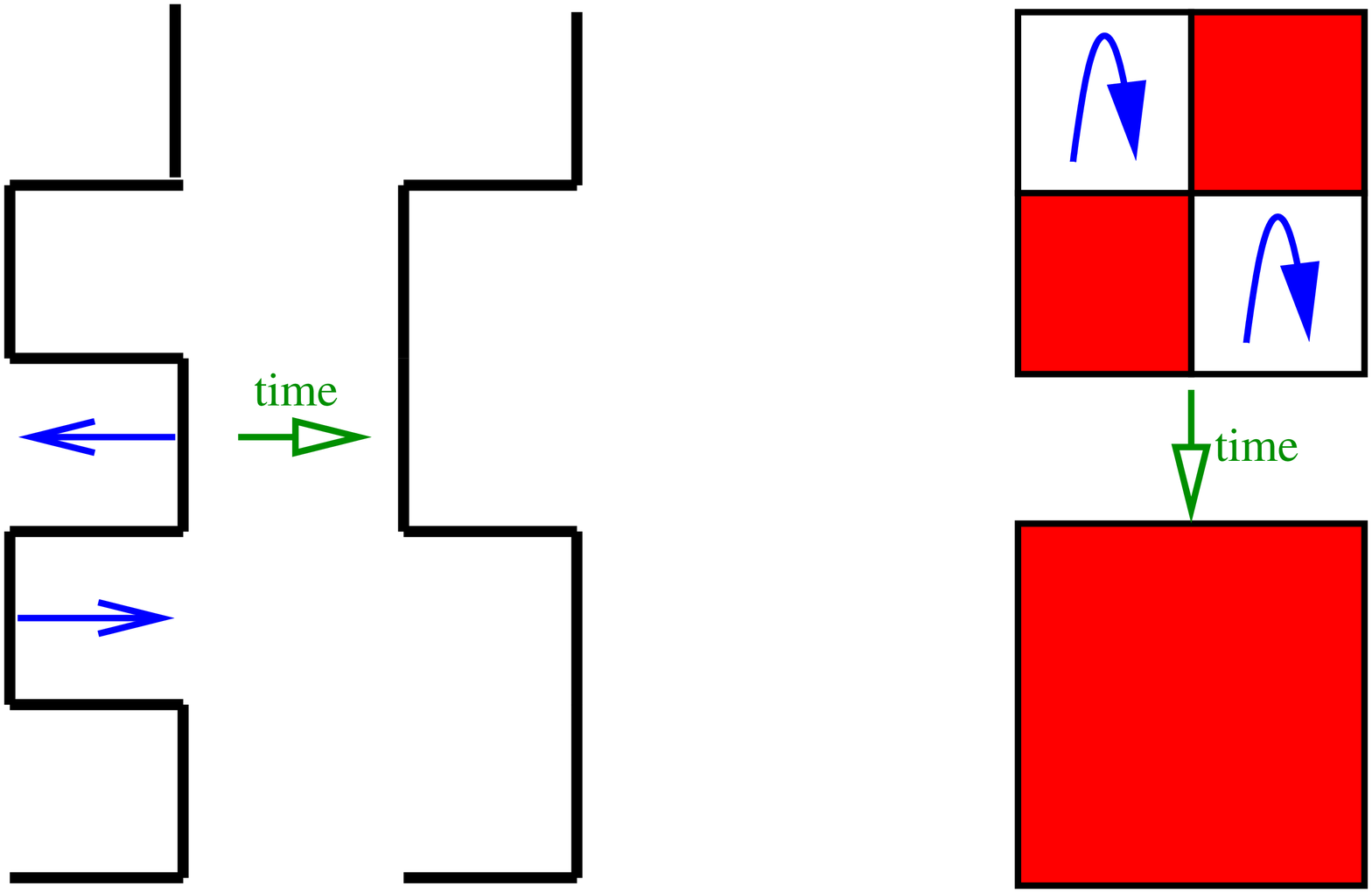}}}
    \caption{(Color online). Upper part: A flat surface in $d=2$ with a distortion (left) and a 
             {\it bubble } of reversed spins (right). 
	     Lower part: The increase of $L(t)$ in $d=2$ when features such as distortions
             (left) or bubbles (right) are removed.}
\label{figbump}
\vspace{2cm}
\end{figure}

We assume that the growing structure can be thought of as made of
{\it features}, with a faceted geometry. 
Features are distortions of flat interfaces
or bubbles of spins. For the square lattice considered 
in the following, these are schematically drawn in Fig.~\ref{figbump} (upper part)
in $d=2$. 

\subsection{Relation between the relaxation of a feature and the exponent $z$}

In order to have coarsening, features must
be progressively removed~\cite{nota1}, by flipping 
all their spins. Let us define $\tau _l $ as the typical time to complete this process
for a feature of size $l$. 
Our strategy is to relate $\tau _l$ to the growth exponent $z$.
In order to do that, let us notice that
when, after a time $\tau _l$,  
features of size $l$ are removed, the typical scale of the
system is increased by a quantity $\Delta l\propto l$, 
as shown in Fig.~\ref{figbump} (lower part).
Assuming scaling, namely the presence of a single relevant lengthscale,
the typical size $l$ of a feature at time $t$ must be of order $L(t)$. 
Therefore $dL(t)/dt\simeq \Delta l /\tau _l$. Let us anticipate what 
will be shown in the next Section, namely that $\tau _l\propto l^\alpha $,
with $\alpha =2$. 
Therefore we have $dL(t)/dt \propto L(t)^{1-\alpha }$,
and so $L(t)\propto t^{1/z}$ with 
\be
z=\alpha =2. 
\label{zalpha}
\ee 

In the following we consider the behavior of $\tau _l$.

\subsection{Relaxation time of a feature} \label{relaxtime}

\begin{figure}
\vspace{2cm}
    \centering
   \rotatebox{0}{\resizebox{.2\textwidth}{!}{\includegraphics{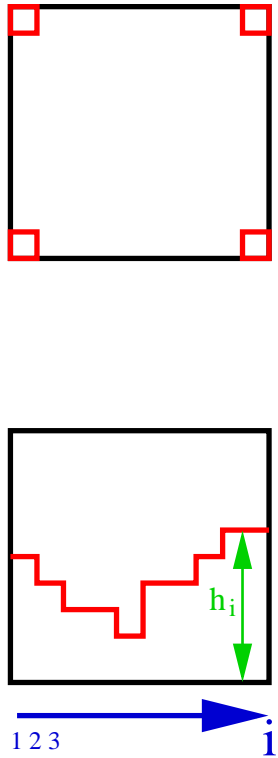}}}
    \caption{(Color online). Relaxation of a bubble in $d=2$. Upper: only the spins in the
     corner can be flipped initially. Lower: a typical configuration
     at a generic time.}
\label{figbump2}
\vspace{2cm}
\end{figure}

We use the terminology of the case $d=2$, for simplicity,
but the argument is general.
Let us consider the relaxation of an initially (at time $t=0$) squared 
bubble, represented in Fig.~\ref{figbump2}.
At zero temperature only spins with $n \le 0$ can be flipped.
Therefore, referring to the situation of the upper part of 
Fig.~\ref{figbump2}, the first move 
is necessarily the flip of one of the four spins in the corners
of the square. 
These moves trigger
a sequence of successive flips, producing the shrinking
of the bubble. Let us suppose, in order to
simplify the argument, that
spins are flipped starting from the bottom of the box
(actually the flipping of the spins proceeds on the average
from each side, but this does not change our conclusions).
Let us denote with $h_i$ ($i=1,2,\dots,l$) the height 
of the $i$-th column of the box at time 
$t$, as shown in Fig.~\ref{figbump2}. Due to the zero temperature
constraints, while the first and last
column, with $i=1$ and $i=l$, are always allowed to grow, 
due to the presence of the wall, all the other columns can
do it only if at least one the nearest
columns is higher. Moreover, a column cannot decrease its height
if it is not higher than at least one of the neighborings.
With these rules, columns evolve until, at $t=\tau _l$, all the 
spins in the box are flipped, the bubble disappears, and the process ends. 

\begin{figure}
\vspace{2cm}
    \centering
   \rotatebox{0}{\resizebox{.5\textwidth}{!}{\includegraphics{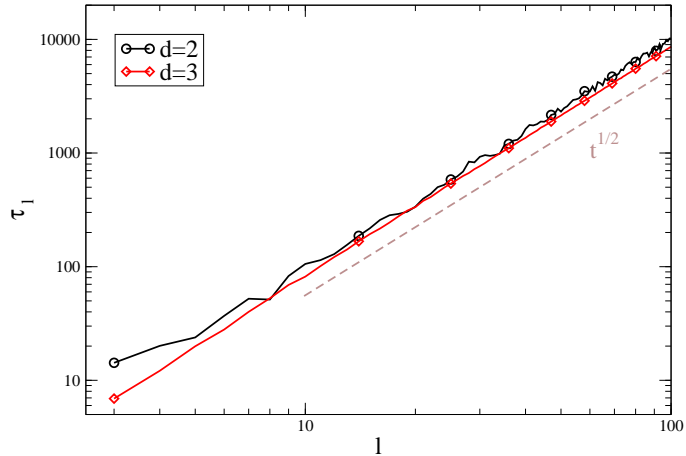}}}
    \caption{(Color online). The typical time $\tau _l$ needed to flip all the spins 
            of a squared bubble of size $l$.}
\label{simulbox}
\vspace{2cm}
\end{figure}

Since with this dynamics an exact evaluation of $\tau _l$ is not possible,
in the following we consider a slightly modified kinetics for
which a determination of $\tau _l$ is allowed; we will then argue,
checking this hypothesis numerically, that
the modification of the dynamics does not change
significantly the behavior of $\tau _l$ and, in particular, the
exponent $\alpha $.  
More precisely, we modify the original dynamics by introducing
an additional constraint, namely $\vert h_{i+1}-h_i\vert \leq 1$.
With this modification the problem can be mapped
onto a diffusion equation for the variables $h_i$.
This result, which applies to the case $d=3$ as well, is
shown in Appendix I. For an interface described by a diffusion equation      
one has $\tau _l \propto l^\alpha $, with $\alpha =2$. 
We argue that the same result applies to the original dynamics
as well. The reason is the following:
due to all the constraints discussed above, the heights $h_i$
are not independent, the typical differences 
$\vert h_{i+1}-h_i\vert $
do not grow very large and, for large $l$, they are independent of $l$. 
This is confirmed by looking at a simulation of the bubble shrinking.
For large $l$, since the differences $\vert h_{i+1}-h_i\vert $ are small
as compared to the relevant scale $l$, we expect that the effect of the 
additional constraint
does not change the exponent $\alpha $. This last statement is convincingly confirmed
by the results of numerical simulations, shown in Fig.~\ref{simulbox}.
This figure refers to the simulation of a squared (cubic in $d=3$) bubble, 
namely an Ising model on an $l^2$ ($l^3$ in $d=3$) squared lattice with
(say) up spins on the boundary and an initial condition of
down spins in the interior. Averaging over several ($10^3-10^5$, depending on $l$) 
realizations of the thermal history, for each value of $l$ we have computed
$\tau _l$ as the time needed to revert the last spin.  Fig.~\ref{simulbox}
shows that $\tau _l \propto l^\alpha$, with $\alpha =2$ is found
with very good accuracy, regardless of dimensionality. This result
confirms the validity of the hypothesis according to which the exponent
$\alpha $ is the same ($\alpha =2$) in the original Ising dynamics and in the modified kinetics considered in Appendix I. 
 
With this result Eq.~(\ref{zalpha}) follows, namely $z=2$ in every
dimension. The exponent $z$ is therefore the same as in quenches 
to finite temperatures. 
Assuming scaling, the size of domains $L(t)$ is related
by $L(t)\propto \rho (t)^{-1}$ to the total density of interfaces
present in the system. The exponent $z$ therefore gives informations
on the number of interfaces, not on their geometry.
In order to appreciate geometrical properties we will
consider in the following other observables. 

\subsection{Densities of spins with a given degree of alignment.} \label{densities}

\begin{figure}
\vspace{2cm}
    \centering
   \rotatebox{0}{\resizebox{.5\textwidth}{!}{\includegraphics{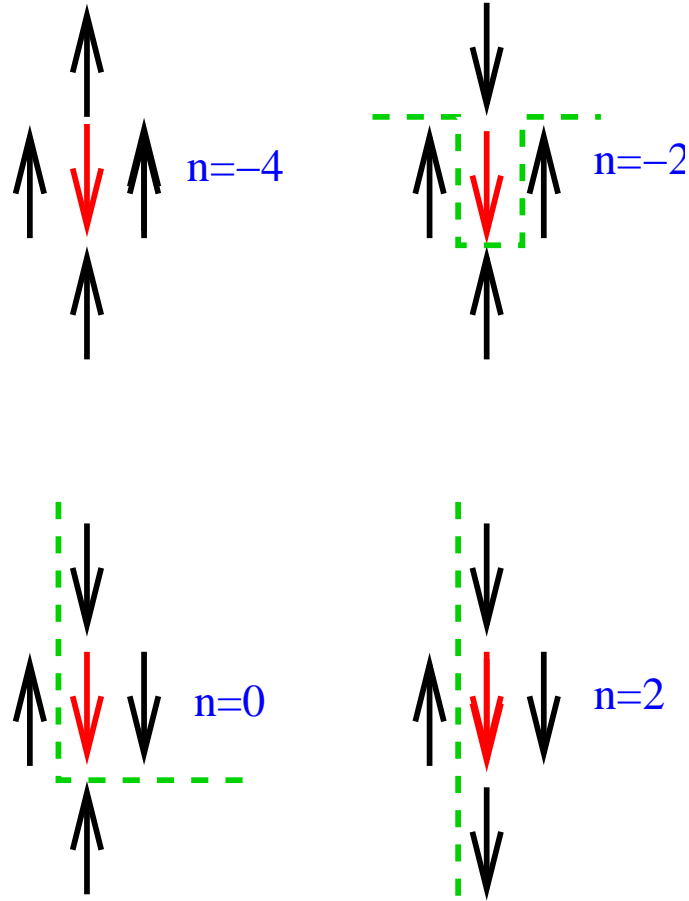}}}
    \caption{(Color online). Classification of interfacial spins in $d=2$. The central spin
    is classified according to the degree of alignment. The
    dashed line is the typical shape of the interface associated
    to each type of spin during the bubble shrinkage.}
\label{class_spins}
\vspace{2cm}
\end{figure}

Restricting again to the $d=2$ case for simplicity (
the extension to the case $d=3$ is straightforward and will be discussed in Appendix II).
we introduce the density $\rho _n(l)$ of spins with
a certain degree of alignment $n_i=n$ in a feature. 
In the following, we will only refer to interfacial spins: bulk
spins with $n=4$, whose behavior is trivial, will never be considered.
Initially, in the squared bubble interfacial spins are those
on the flat boundaries or in the corners, 
with $n=2$ and $n=0$ respectively.
During the evolution, 
as shown in the lower part of Fig.~\ref{figbump2}, all the possible
values of $n$ can be generated. The set of $\rho _n (l)$ provides 
a geometric characterization of the interface.
A representation of the typical geometry where a spin with
a degree of alignment $n$ occurs when the bubble shrinks
is drawn in Fig.~\ref{class_spins} 

While columns are growing, a generic profile of $h_i$ is made of
steps, namely spins with $n=0$,  and flat parts
with $n=2$, as shown in Fig.~\ref{figbump2}.
A step on site $i$ can be randomly replaced by a flat part
and the reverse is possible as well. As a consequence 
for sufficiently large values of $l$ a finite fraction,
independent of $l$, of spins 
with $n=0$ and $n=2$ will be typically present. Simulations
clearly show that these numbers, on average,
do not depend on time (excluding, possibly, the initial
and final stages of the process). 
According to this,  the number of spins with $n=0$ or $n=2$ is constant and
proportional to the length of the interface, namely to $l$.
Normalizing with the total number of spins $l^2$, we obtain
$\rho _2(l)\propto l^{-1}$ and $\rho _0(l)\propto l^{-1}$.
The situation is very different for spins with $n=-2$ and $n=-4$.
The former can only be produced when a last spin must be reversed in order 
to complete a row, an event happening, on average, every $l$ moves. 
When this occurs, a single spin (out of $l^2$) with $n=-2$ is generated.
Looking at a generic time, therefore, the typical density of such
spins is $\rho _{-2} (l)\propto l^{-3}$.
Finally, spins with $n =-4$ are only obtained 
when the last spin of the box must be reversed. 
In a bubble of $l^2$ spins only one can be the last,
and this happens once every $l^2$ moves.
Hence $\rho _{-4} (l)\propto l^{-4}$. Again, scaling implies that
we can identify $l$ and $L(t)$, leading to 
\be
\rho _n [L(t)]\propto L(t)^{-\beta_n},
\ee
with 
\be
\beta _2=\beta _0=1,
\label{alfaf}
\ee
\be
\beta _{-2}=3,
\ee
\be
\beta _{-4}=4.
\label{alfai}
\ee
This argument can be extended to the case $d=3$ (see Appendix II).
The results are 
\be
\beta _4=\beta _2=\beta _0=1,
\label{alfaf3d}
\ee
\be
\beta _{-2}=3,
\ee
\be
\beta _{-4}=\beta _{-6}=4.
\label{alfai3d}
\ee
In the next Section we will compare these predictions with the outcome of 
numerical simulations, both in $d=2$ and $d=3$.

\section{Numerical simulations} \label{numerical}

\subsection{Quenches to $T_f=0$.}\label{numteq0}

\begin{figure}
\vspace{2cm}
    \centering
   \rotatebox{0}{\resizebox{.5\textwidth}{!}{\includegraphics{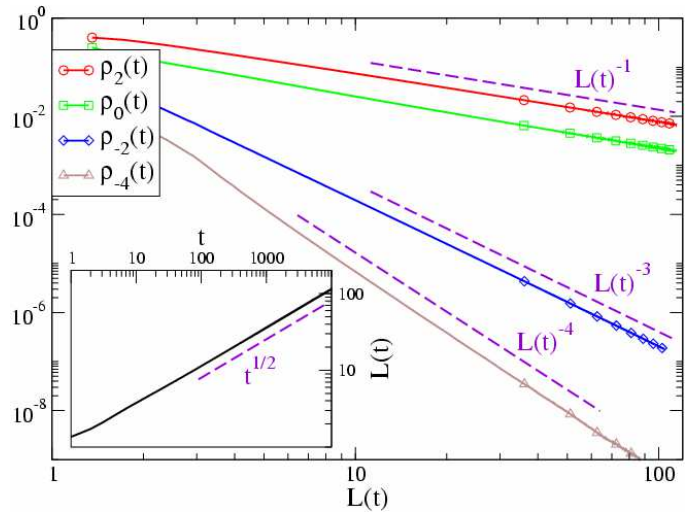}}}
    \caption{(Color online). Quench to $T_f=0$ in $d=2$. 
      The densities $\rho _n [L(t)]$ are plotted against $L(t)=\rho (t)^{-1}$. 
      In the inset
             $L(t)$ is plotted against time.}
\label{quencht02d}
\vspace{2cm}
\end{figure}

We have simulated systems of $2000^2$ and $576^3$ spins in $d=2$ and $d=3$,
respectively, on square (cubic) lattices. With these sizes, we have checked that finite size effects are not 
present in the range of times presented in the figures.
The critical temperature of the model is $T_c\simeq 2.269$ and
$T_c\simeq 4.512$ in $d=2$ and $d=3$ respectively (we set $J=1$).
Time is measured in montecarlo steps (mcs).
In Fig.~\ref{quencht02d} we show the results for $d=2$. 
Here we observe that a scaling regime, attested by the power law
behavior of all the plotted quantities, sets in after a very short time
$t_{sc}\simeq 4$ mcs. Best power law fits to the data (for $t\ge 10$)
give $\beta _{2}=0.99\pm 0.02$, $\beta _{0}=1.03\pm 0.03$, $\beta _{-2}=2.99\pm 0.04$, 
$\beta _{-4}=4.05\pm 0.05$, and $z=1.99\pm 0.02$.
All the exponents $\beta _n$ and $z$ are in excellent agreement with
the determination made in the previous Section. 

When scaling
holds, according
to Eqs.~(\ref{1.132}),(\ref{1.1321}) the equal time correlation function,  
behaves as
\be
G_{ag}(r,t,t)=h(r/L(t)),
\label{acca}
\ee
with $h$ obeying Porod law~(\ref{porod})
in the case of sharp interfaces. In the case considered here,
$G_{ag}(r,t,t)=G(r,t,t)$, since $G_{st}(r,t,t)\equiv 0$ at $T_f=0$.  
According to Eq.~(\ref{acca}), when scaling holds curves of 
$G_{ag}(\vec r,t,t)$ for different times should collapse when plotted against $r/L(t)$. 
This is observed in Fig.~\ref{figporod} (left part). Besides, the Porod
law~(\ref{porod}) is very neatly obeyed. 

\begin{figure}
\vspace{2cm}
    \centering
   \rotatebox{0}{\resizebox{.5\textwidth}{!}{\includegraphics{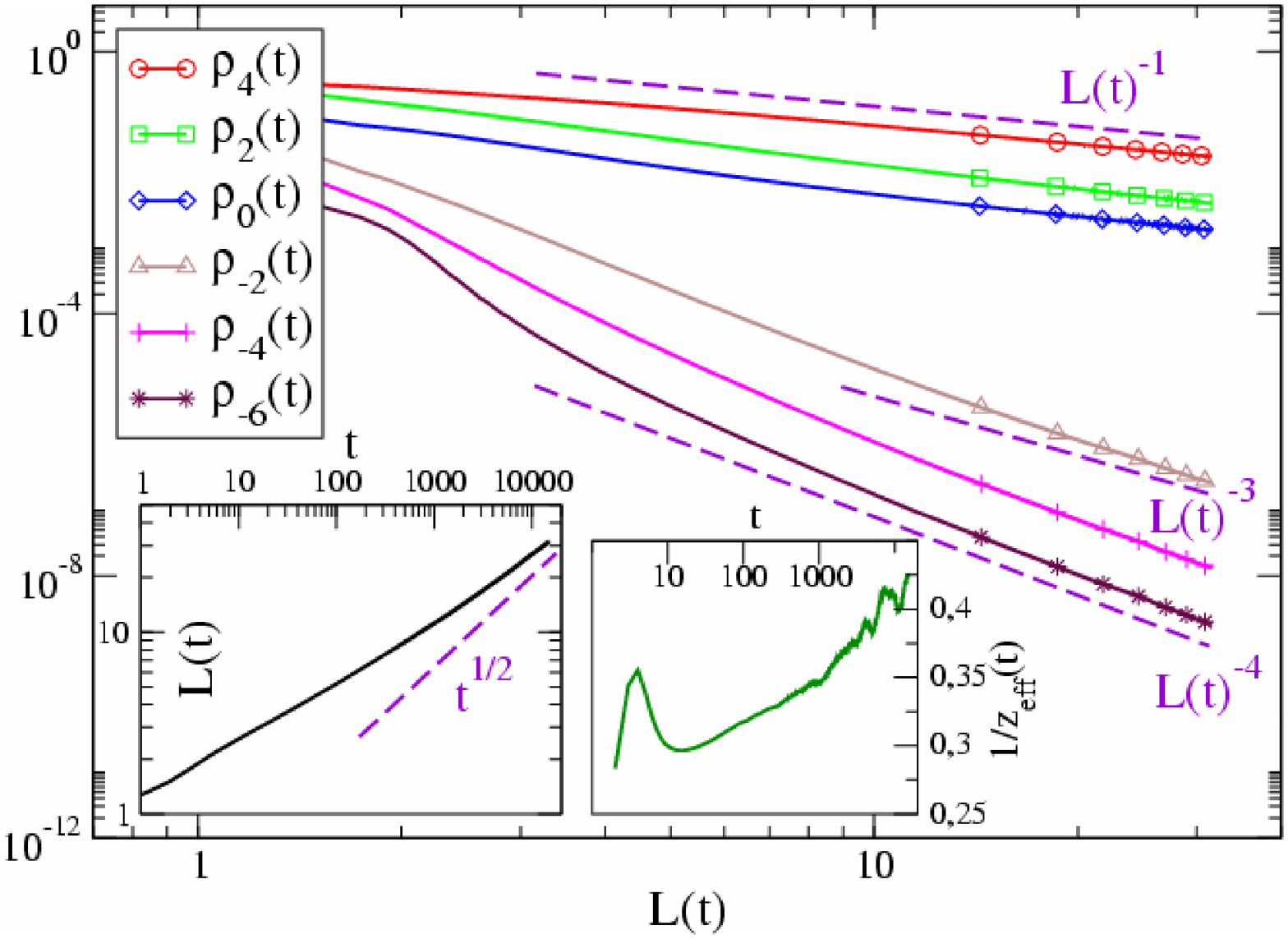}}}
    \caption{(Color online). Quench to $T_f=0$ in $d=3$. The densities 
      $\rho _n [L(t)]$ are plotted against $L(t)=\rho (t)^{-1}$. In the insets
             $L(t)$ and $1/z_{eff}(t)$ are plotted against time.}
\label{quencht03d}
\vspace{2cm}
\end{figure}

In Fig.~(\ref{quencht03d}) a plot analogous to that of 
Fig.~\ref{quencht02d} is made for $d=3$.
After a time around 10 mcs a power-law behavior sets in for all the 
$\rho _n [L(t)]$. Fitting the curves with power laws
we find a residual time dependence of the exponents $\beta _{n}$,
since their value changes measuring them in different timewindows.
This is particularly evident for $\beta _{-2}$.
This indicates that preasymptotic corrections to scaling
are not completely negligeable in the timedomain of our
simulations.
Best power-law fits for $t>10$ yield 
$\beta _{4}=0.97\pm 0.04$, $\beta _{2}=1.11\pm 0.06$, $\beta _{0}=1.03\pm 0.04$,
$\beta _{-2}=3.3 \pm 0.1$, $\beta _{-4}=3.8\pm 0.1$, and $\beta _{-6}=3.9\pm 0.1$.
Taking into account the presence of preasymptotic corrections
we regard these values as being
consistent with the results of the previous Section. 
Regarding $L(t)$, instead, the data (in the inset)
do not show a satisfactory power law. The curve is bending upwards on the
double logarithmic plot and the exponent is not consistent with the expected
value $z=2$.
In order to clarify this point we have computed the effective exponent 
\be
\frac{1}{z_{eff}(t)}= \frac{d\ln L(t)}{d\ln t},
\label{effexp}
\ee
which is shown in the inset of Fig.~\ref{quencht03d}.
In an early stage, when scaling does not hold, this quantity
grows exponentially as described by linear theories~\cite{Cahn}.
Then, after reaching a minimum of order $0.3$, $1/z_{eff}(t)$ it keeps slowly,
but steadily, increasing.
Its value measured at the longest times is around $1/z_{eff}\simeq 0.43$.
Actually, to the best of our knowledge,
the expected exponent $z=2$ has never been reported. Previous simulations
on much shorter timescales observed~\cite{Amar89} an exponent of order $1/3$ 
and sometimes the very existence of
dynamical scaling has been questioned. Notice that the value 
$1/3$ is comparable to the
value $1/z_{eff}\simeq 0.3$ of the effective exponent around its 
minimum, a fact which may explain what reported in~\cite{Amar89}. 
Our data are consistent with a possible asymptotic
value $z=2$ although much larger simulation efforts would be needed
for a definitive evidence. In any case, the data show that preasymptotic
effects are quite relevant, and exclude that a well defined 
exponent can be measured, up to $t\simeq 2\cdot 10^4$ mcs
(for longer times, not reported in the Figure, finite size effects 
are observed). 
A rough extrapolation suggests that times at least
10 times larger ($t\simeq 2\times 10^5$ mcs) are needed to observe 
$1/z_{eff}(t)\simeq 1/z\simeq 1/2$ (meaning a lattice size of order
$(2\cdot 10^3)^3$ in order to be finite size effects free).


\subsection{Quenches to $T_f>0$.}\label{tneq0}

\begin{figure}
\vspace{2cm}
    \centering
   \rotatebox{0}{\resizebox{.5\textwidth}{!}{\includegraphics{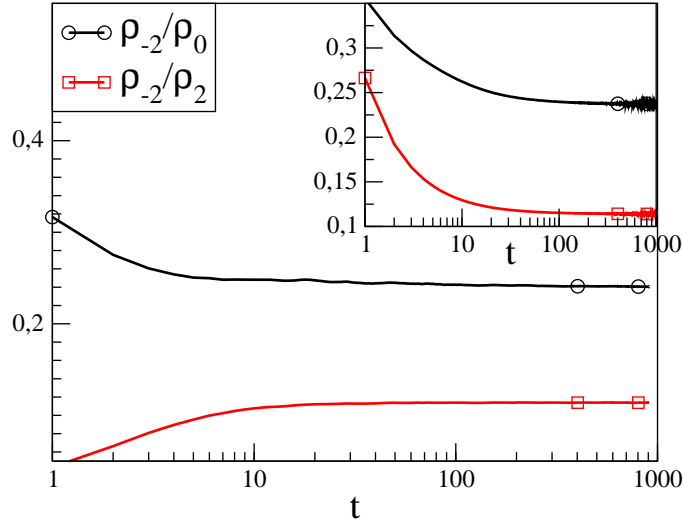}}}
\vspace{1cm}
    \caption{(Color online). 
      The ratios between $\rho _{-2} (l)$, $\rho _0 (l)$, and $\rho _{2} (l)$ 
             are plotted against $t$ 
             for an equilibrium interface in $d=2$ at $T=2$ (main) and
             in a quench to $T_f=2$ (inset). NBF dynamics is used in
             both cases.}
\label{quenchvssingleinterf}
\vspace{2cm}
\end{figure}

\begin{figure}
\vspace{2cm}
    \centering
   \rotatebox{0}{\resizebox{.45\textwidth}{!}{\includegraphics{quencht22d.eps}}}
   \hspace{1cm}
   \rotatebox{0}{\resizebox{.45\textwidth}{!}{\includegraphics{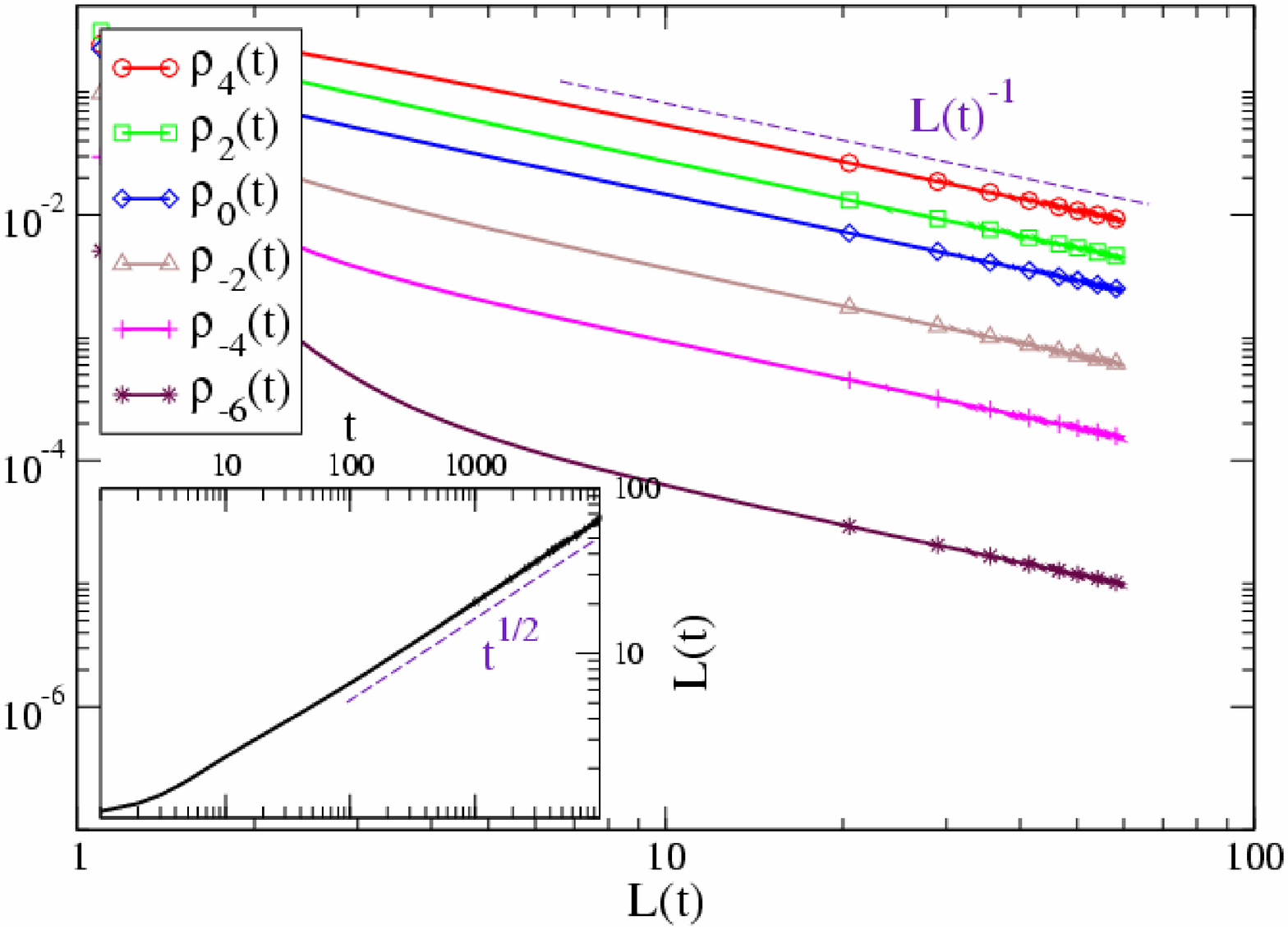}}}
    \caption{(Color online). Quench to $T_f=2$ (NBF), for $d=2$ (left) and $d=3$ (right).
      $\rho _{n} [L(t)]$ are plotted against $L(t)=\rho (t)^{-1}$. In the inset
             $L(t)$ is plotted against time.}
\label{quencht2}
\vspace{2cm}
\end{figure}

When the quench is made to a finite final temperature all the constraints imposed by 
$T_f=0$ are removed. In the quench to $T_f=0$, as seen in Sec.~\ref{numteq0}, 
spins with $n>0$ are present in the system, but they cannot be updated because
this would increase the energy. When $T_f>0$
also these can be updated, although with a small probability 
for small $T_f$.
For shallow quenches ($T\lesssim T_c$) the typical times 
$\tau _{n}(T_f)= W(n_i[\sigma],T_f)^{-1}$ associated
to microscopic moves are small, and, in particular, much smaller then the timescales
over which the non-equilibrium behavior of interfaces takes place. Therefore we expect
that during phase-ordering an interface be in {\it quasi equilibrium}, namely
it will have the same values of $\rho _n [L(t)]$
of an equilibrium interface~\cite{nota4} of length  $l=L(t)$ at 
the same temperature of the quench. 
In order to check this conjecture we have performed the following simulations:
In $d=2$ we have prepared an Ising system with a spanning vertical interface in the middle 
and antiperiodic boundary conditions in the horizontal direction; subsequently, 
we let it evolve at a constant temperature $T$ and, during the
evolution, we have measured the densities $\rho _n (l)$. These quantities
are compared
to the quantities $\rho _n [L(t)]$ measured in a quench to the final temperature $T_f=T$.
In both the situations we have
implemented a fast dynamics where spins with $n=4$
are not allowed to flip. 
This no bulk flip (NBF) dynamics has been frequently used in
the literature~\cite{Corberibis,Corberi03}. Apart from its numerical efficiency, it has the
advantage of isolating the aging behavior of the system.
The reason is that since, as already pointed out in Sec.~\ref{intro}, the stationary terms in Eq.~(\ref{split}) are produced by the flipping
of the spins inside the domains, by preventing bulk moves one
is left only with the dynamics of the interfaces which is responsible
for the aging term of Eq.~(\ref{split}). On the other hand, it has
shown~\cite{Corberi03} also that the NBF rule does not change
the properties of the aging terms in the large time domain.   
In the simulation of the single interface we use this no bulk flip (NBF) 
dynamics also as a tool to
maintain a single interface in the system at all times. With the standard dynamics
spins can be reversed in the bulk, creating additional interfaces that may interact
with the original spanning interface. On the other hand, with NBF dynamics,
the spanning interface remains unique and well defined at all times. 
In order to avoid the complications arising when comparing $\rho _n (l)$ with 
$\rho _n [L(t)]$ in the two kind of simulations, which would require the comparison of
the size $l$ of the interface in the equilibrium simulation 
with the length $L(t)$ in the corresponding quenched system, in Fig.\ref{quenchvssingleinterf}
we have plotted the ratios between different $\rho _n$. 
Since these quantities do not depend on $l$ or on $L(t)$ respectively in the two cases,
they can be directly compared.   
After a brief transient, 
the single interface (main figure) reaches the stationary state and the ratios between 
different $\rho _n $ take time-independent values.
The same is true for the quenched system, shown in the inset.
We find that the asymptotic value of the ratios is the same
with good accuracy in the two systems. This confirms our claim that in a shallow quench 
the values of the densities $\rho _n [L(t)]$ are equal to the corresponding
quantities $\rho _n (l)$ in an equilibrium interface of size $l$. 
Since the latter are finite and time-independent
this implies that in a shallow quench $\rho _n [L(t)]\propto \rho (t)$. 
Therefore, instead of Eqs.~(\ref{alfaf}-\ref{alfai}), and Eqs.~(\ref{alfaf3d}-\ref{alfai3d}) 
one must have
\be
\beta _n=1,
\label{betaeq1}
\ee
for every value of $n$.
This is shown to be true in Fig.~\ref{quencht2} for systems in $d=2,3$ quenched to $T_f=2$. 
Best power-law fits (for $t>10$) yield $\beta_{2}=1.00 \pm 0.01$, 
$\beta_{0}=1.00 \pm 0.01$,
$\beta_{-2}=1.00 \pm 0.01$, $\beta_{-4}=1.01 \pm 0.1$, in $d=2$ and
$\beta_{4}=1.00 \pm 0.01$,$\beta_{2}=1.00 \pm 0.01$,$\beta_{0}=1.00 \pm 0.01$,
$\beta_{-2}=1.00 \pm 0.01$,$\beta_{-4}=1.01 \pm 0.01$, $\beta_{2}=1.01 \pm 0.02$,
in $d=3$.
Notice that the values of $\beta _{n}$ with $n<0$ are very different
from the case with $T_f=0$ and can be used to distinguish the two kinds of dynamics.
Regarding the value of the exponent $z$, the curvature driven
mechanism implies $z=2$. Actually, this is found with very good accuracy
in $d=3$ (we find $1/z=0.502 \pm 0.04$ in the range $t\in [10^2-10^4]$).
Instead, for $d=2$ 
one observes a slightly larger exponent, since $1/z$ is of order 0.48
in the region of the largest simulated times. 
We will comment later on this point. 

\begin{figure}
\vspace{2cm}
    \centering
   \rotatebox{0}{\resizebox{.40\textwidth}{!}{\includegraphics{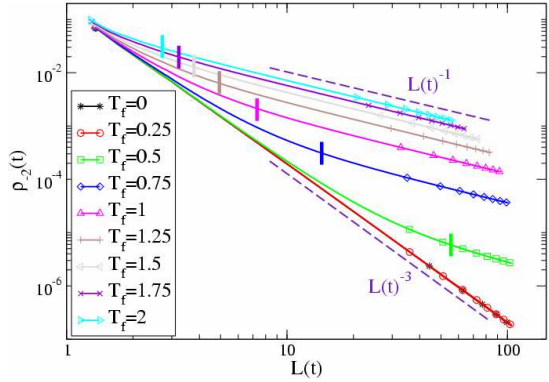}}}
   \hspace{1cm}
   \rotatebox{0}{\resizebox{.40\textwidth}{!}{\includegraphics{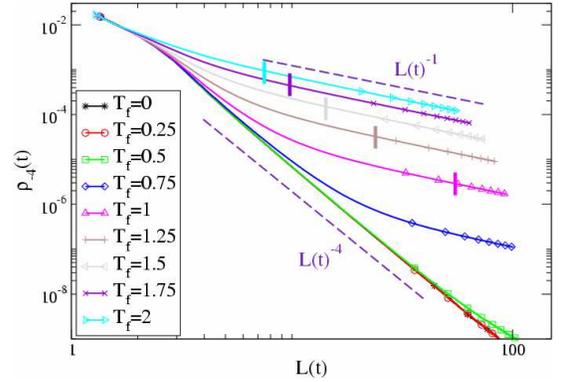}}}
    \caption{(Color online). $d=2$. $\rho _{-2}[L(t)]$ (left) and $\rho _{-4}[L(t)]$ (right) 
             are plotted against $L(t)=\rho (t)^{-1}$ (NBF). Vertical segments 
             across the curves represent the crossover lengths $L_{-2}(T_f)$ and
             $L_{-4}(T_f)$ of Eq.~(\ref{crosslength}) (when reached in the simulation).}
\label{crossover}
\vspace{2cm}
\end{figure}

\begin{figure}
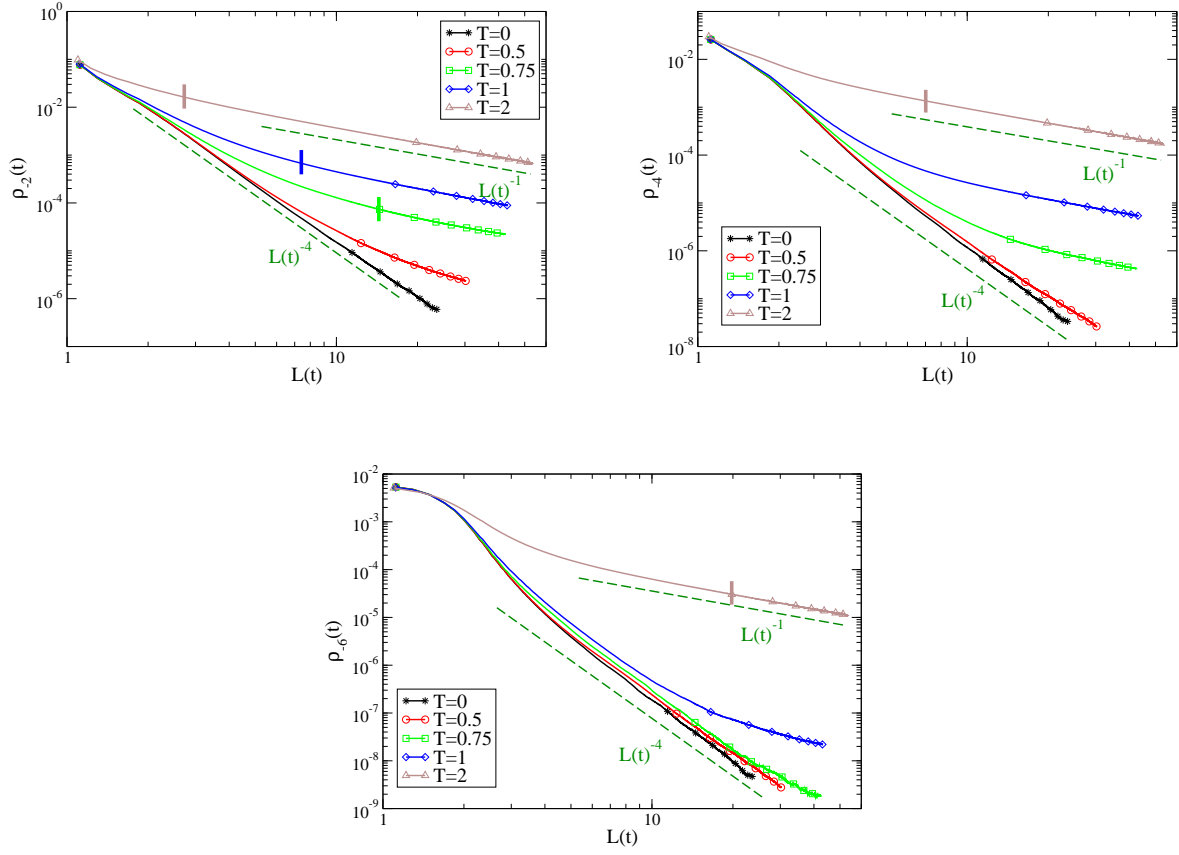

\vspace{2cm}
    \centering
   \rotatebox{0}{\resizebox{.40\textwidth}{!}{\includegraphics{cross1_3d.eps}}}
\vspace{1cm}
\hspace{1cm}
   \rotatebox{0}{\resizebox{.40\textwidth}{!}{\includegraphics{cross2_3d.eps}}}
\vspace{1cm}
   \rotatebox{0}{\resizebox{.40\textwidth}{!}{\includegraphics{cross3_3d.eps}}}
    \caption{(Color online). $d=3$. 
            $\rho _{-2}[L(t)]$, $\rho _{-4}[L(t)]$ and $\rho _{-6}[L(t)]$ 
             are plotted against $L(t)=\rho (t)^{-1}$ (NBF). Vertical segments 
             across the curves represent the crossover lengths 
	     $L_{-2}(T_f)$, $L_{-4}(T_f)$ and $L_{-6}(T_f)$ of Eq.~(\ref{crosslength}) 
(when reached in the simulation).}
\label{crossover3d}
\vspace{2cm}
\end{figure}

Since the dynamics is different in shallow quenches or in quenches to 
$T_f=0$ we expect to see a crossover phenomenon
at intermediate temperatures. Namely, for every 
$T_f>0$ a crossover time should exist separating an early stage where the dynamics
is of the $T_f=0$ type, with $\beta _n$ given in Eqs.~(\ref{alfaf}-\ref{alfai})
for $d=2$ or in Eqs.~(\ref{alfaf3d}-\ref{alfai3d}) for $d=3$, 
from a late stage where the finite temperature scalings~(\ref{betaeq1})
set in. For a class of spins with a given degree of alignment $n$ the crossover 
between the early and the late kind of dynamics occurs
when spins with the considered $n$ start to be created by means of
activated moves. The crossover time, therefore, should be 
of order
$\tau _n(T_f)\simeq W(n,T_f)^{-1}$, and is therefore different
for spins with different $n$. At the crossover time $\tau _n(T_f)$
a typical crossover length 
\be
L_n(T_f)\simeq \tau _n(T_f)^{1/z}=W(n,T_f)^{-1/z}
\label{crosslength}
\ee 
is associated.
In Figs.~\ref{crossover},\ref{crossover3d}, the pattern of crossover described above
can be observed. Here we see that, practically for all the $T_f$ considered, 
the behavior of the densities $\rho _n(t)$ is initially analogous to that
of the $T_f=0$ case (Eqs.~(\ref{alfaf}-\ref{alfai})
or Eqs.~(\ref{alfaf3d}-\ref{alfai3d}) in $d=2$ or $d=3$ respectively).
This regime lasts until $L(t)\simeq L_n(T_f)$, where 
$\rho _n(t)$ start to behave as in shallow quenches (Eq.~\ref{betaeq1}).
For very small $T_f$, $L_n(T_f)$ is outside the range of simulated times
(this explains why, for instance, the curves with $T_f=0$ and $T_f=0.25$
can be hardly distinguished in $d=2$). Increasing $T_f$ gradually, $L_n(T_f)$,
whose values obtained from Eq.~(\ref{crosslength}) are marked with vertical 
segments across the curves (when within the
simulated times), become progressively smaller. One observes that the
crossover phenomenon occurs at different times for spins with different 
$n$, and that the estimate~(\ref{crosslength}) agrees reasonably well
with what observed.

\begin{figure}
\vspace{2cm}
    \centering
   \rotatebox{0}{\resizebox{.8\textwidth}{!}{\includegraphics{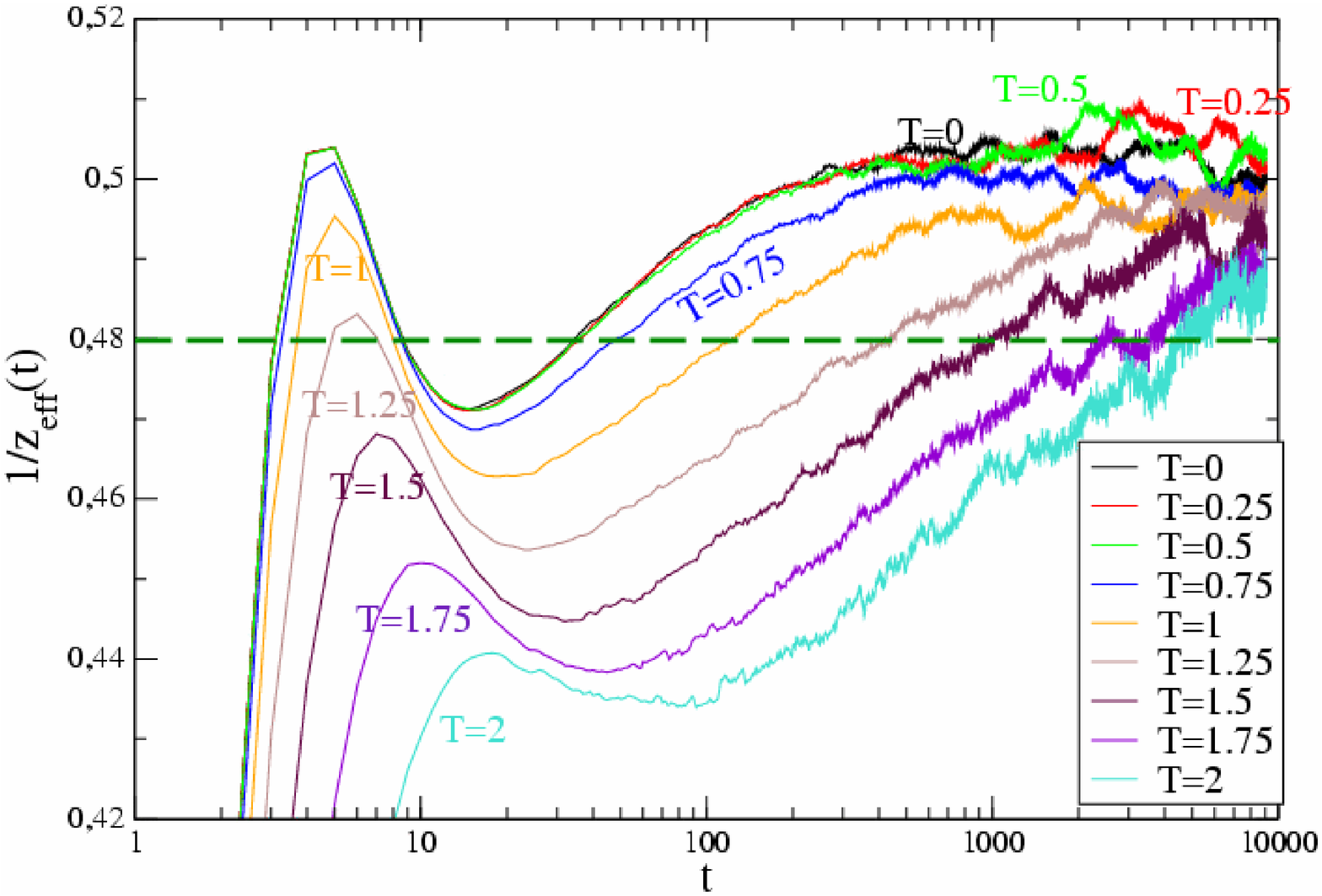}}}
    \caption{(Color online). The effective exponent $1/z_{eff}(t)$ for quenches in $d=2$ (NBF).}
\label{effexp2d}
\vspace{2cm}
\end{figure}

\begin{figure}
\vspace{2cm}
    \centering
   \rotatebox{0}{\resizebox{.8\textwidth}{!}{\includegraphics{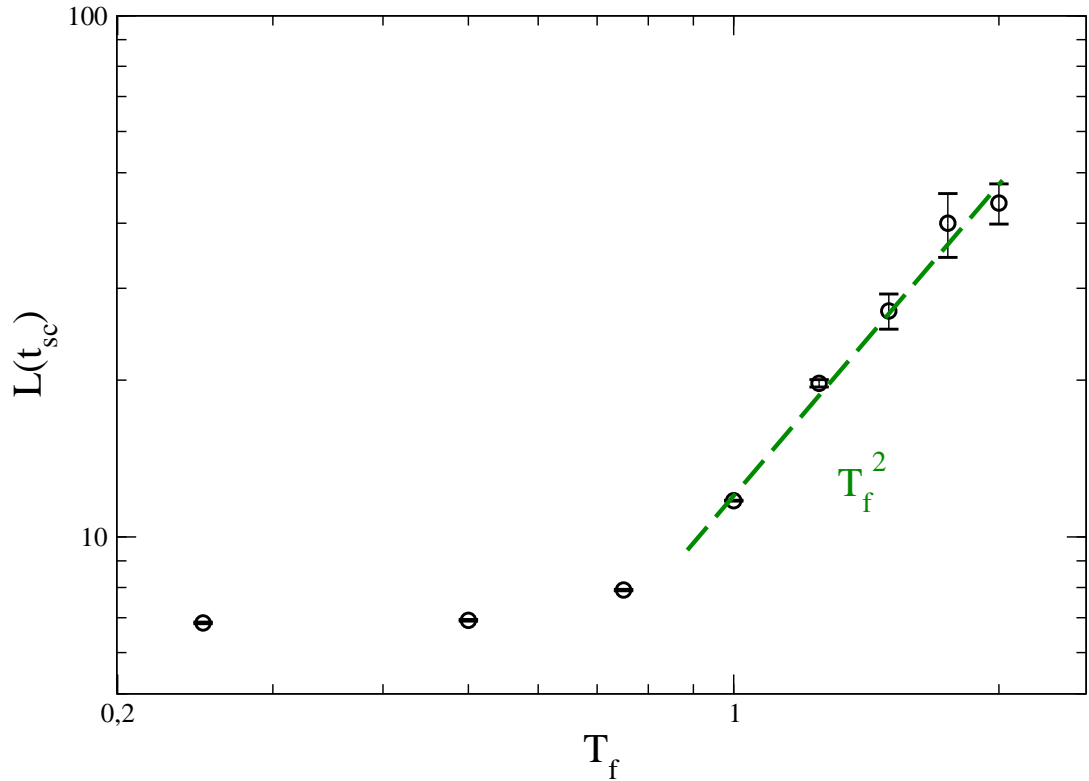}}}
    \caption{(Color online). $L(t_{sc})$ is plotted against $T_f$.}
\label{lscal}
\vspace{2cm}
\end{figure}

Let us now come back to the value of the exponent $1/z$ in $d=2$, which, as already
observed regarding Fig.~\ref{quencht2}, has a value slightly smaller than
the expected one $1/z=0.5$. In order to make more precise statements
we have measured the effective exponent, which is shown in Fig.~\ref{effexp2d}
for various temperatures. For $T=0$ the effective exponent initially
rise to a maximum for the reasons already discussed for the case with $d=3$. Then it goes down to  
a minimum and, later, reaches the
asymptotic value $0.5$ already at times of order $t\simeq 300$ mcs. 
As $T_f$ is increased the pattern is similar but the initial minimum is 
depressed and delayed so that for the largest temperatures considered
$1/z_{eff}(t)$ has not yet reached the asymptotic value 
at the longest simulated times.
Although the expected final value $1/z_{eff}=1/z=1/2$ is not in doubt,
a rough determination of this exponent in 
a simulation may lead to a smaller value, as sometimes reported~\cite{Manoy00}.  
This behavior can be interpreted as due to the presence, 
beside $L(t)$, of another length,
the roughness of the interfaces. Equilibrium interfaces
are rough for $T>T_R$. The roughening temperature $T_R$ vanishes for $d=2$
while $0<T_R<T_c$ for $d=3$.
An interface spanning a box of linear size $l$ in equilibrium
at the temperature $T$ has a typical width $u_{l}(T)$ given by~\cite{Barabasi}
\be
    u_l(T) = \left \{ \begin{array}{ll}
	a_2(T)\sqrt l   \qquad $for$ \qquad d =2 \\
	a_3(T)        \qquad $for$ \qquad d =3, T\leq T_R \\
        a_3 (T)\ln l  \qquad $for$ \qquad d =3, T>T_R.
        \end{array}
        \right .
        \label{wd}
\ee
In the phase-ordering kinetics it has been conjectured
by Villain~\cite{Abraham89} that
the role of $l$ in Eq.~(\ref{wd}) is played by $L(t)$. The 
non-equilibrium width $U(t,T_f)$ should than behave as
\be
    U(t,T_f) \propto \left \{ \begin{array}{ll}
	a_2(T_f)\sqrt {L(t)}   \qquad $for$ \qquad d =2 \\
	a_3(T_f)        \qquad $for$ \qquad d =3, T_f\leq T_R \\
        a_3 (T_f)\ln {L(t)} \qquad $for$ \qquad d =3, T_f>T_R.
        \end{array}
        \right .
        \label{wdoff}
\ee
According to these expressions, in the large time limit $U(t,T_f)$ can always
be neglected with respect to $L(t)$. However, there can be
an initial regime, for $t< t_{sc}$, where $U(t,T_f)$
produces a correction to scaling. In this range of times we expect a 
(time dependent) effective 
exponent $z_{eff}(t)\neq 2$ to be observed.
Given the behaviors~(\ref{wdoff}),
$t_{sc}$ may be sufficiently large to produce observable effects for $d=2$ while
we expect it to be too small to significantly affect the scaling 
behavior for $d=3$.
Actually, we have already observed (see Fig.~\ref{quencht2}) that, 
differently from $d=2$, in $d=3$ 
the effective exponent quickly
converges to the value $z=2$ for quenches to $0<T_f<T_c$.
According to our hypothesis, since $a(T_f)$ 
is an increasing function of $T_f$, while $L(t)$ is roughly temperature independent,
the convergence towards the asymptotic $z=2$
should be delayed increasing $T_f$: This is actually observed in Fig.~\ref{effexp2d}.
In order to check further the consistency
of this hypothesis we have computed the temperature dependence of $a(l,T_f)$.
From a set of simulations of a single interface as those described above
in this section we have extracted the equilibrium width of the interface as
$u_l(T)=\langle 
\sqrt {\sum _{j=1}^l (1/l)[x_j(t)-l/2]^2}\rangle$,
where $j$ is the vertical coordinate in the simulation box,
$x_j(t)$ is horizontal coordinate of the interface position at a generic time $t$, 
and $\langle \cdots \rangle$
is an average over thermal realizations.
Since in equilibrium $u_l(T)$ does not depend on time 
we have also averaged $u_l(T)$ over time in order to reduce the noise.
\begin{figure}
\vspace{2cm}
    \centering
   \rotatebox{0}{\resizebox{.8\textwidth}{!}{\includegraphics{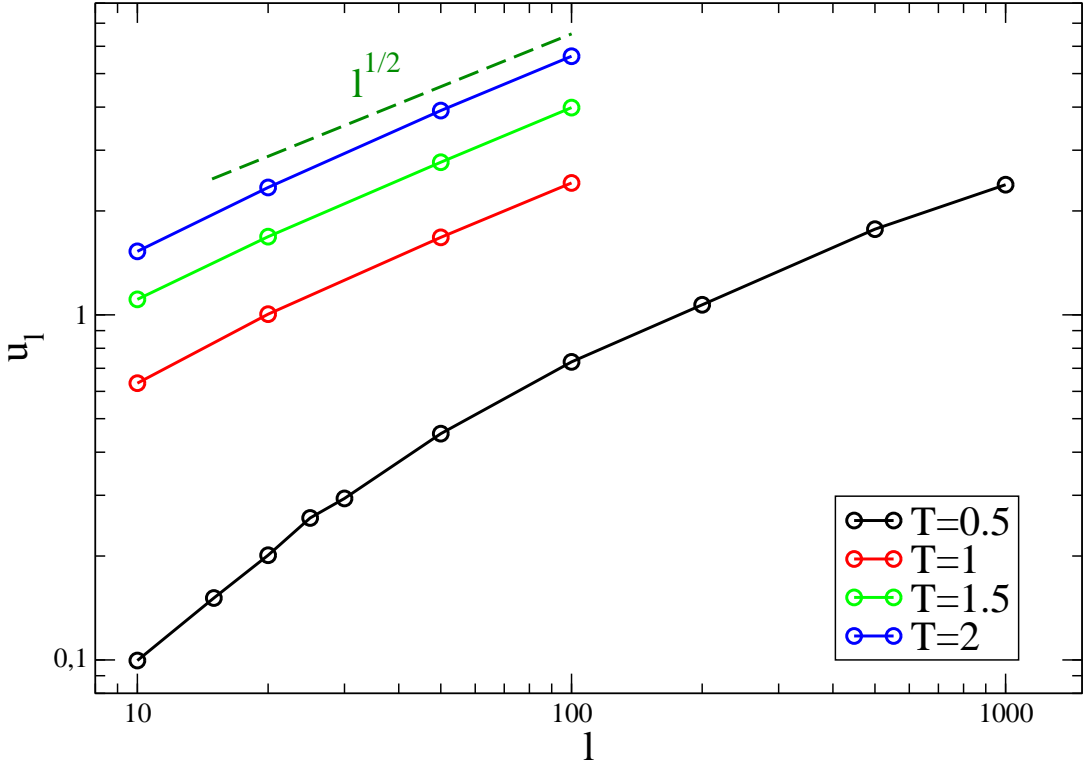}}}
    \caption{(Color online). $u_l(T)$ is plotted against $l$ for different temperatures.}
\label{figwidth1}
\vspace{2cm}
\end{figure}
Fig.~\ref{figwidth1} shows that the behavior~(\ref{wd}) 
is obeyed for $l$ sufficiently large (the larger the lower is $T_f$).
Extracting $a_2(T_f)$ we find a linear relation 
\be
a_2(T_f)=A T_f,
\label{depat}
\ee
where $A$ is a constant ($A\simeq 0.32$).  
We can evaluate $t_{sc}$ from the condition $L(t_{sc})= U(t_{sc},T_f)$.
In a quench from high temperature $L(t)$ start growing from
an initial value $L(0)\simeq 1$. 
For low temperatures, since $a_2(T_f)$ is very small, $L(0)$ is  
larger than $U(0,T_f)$ and hence $U(t,T_f)$ is negligible from the
beginning. In this case scaling can set in very early,
after the microscopic time $t^*\simeq 1$ necessary for the formation
of domains of the equilibrium phases which is practically independent of $T_f$. 
For larger temperatures there is a transient during which
$U(t,T_f)$ cannot be neglected.
Using Eqs.~(\ref{wdoff},\ref{depat}) one obtains $L(t_{sc})=A^2T_f^2$.
Since $L(t)\sim t^{1/2}$ is roughly obeyed also for $t<t_{sc}$ (the effective exponent
is always in the range $[0.45-0.5]$) one can estimate $t_{sc}\propto T^4$.
In Fig.~\ref{lscal} we have plotted $L(t_{sc})$ for different values 
of $T_f$. This quantity have been obtained as
follows: From the data of
Fig.~\ref{effexp2d} we have estimated $t_{sc}$ as the time when $1/z_{eff}(t)$ 
reaches the value $0.48$ (clearly, we refer to the asymptotic
increase of $1/z_{eff}(t)$, for $t>100$, not to the early maximum). 
Successively, from the numerical data for
$L(t)$ we have extracted $L(t_{sc})$. The picture shows agreement 
with the prediction of our hypothesis,
namely a constant value of $L(t_{sc})$ at low temperatures and a behavior
$L(t_{sc})\sim T_f^2$ for larger temperatures.  

\begin{figure}
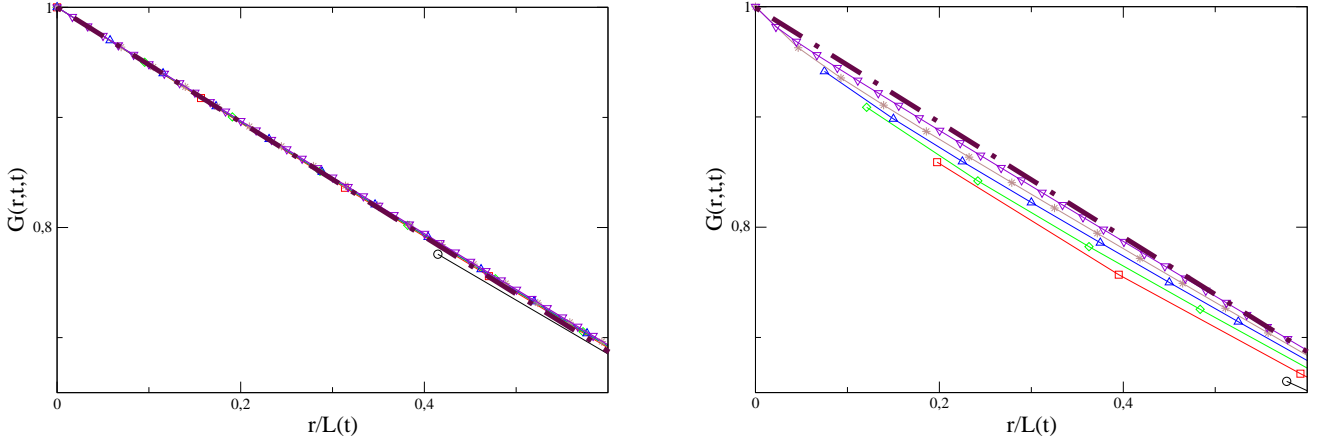

\vspace{2cm}
    \centering
   \rotatebox{0}{\resizebox{.45\textwidth}{!}{\includegraphics{scalingt0.eps}}}
   \hspace{1 cm}
   \rotatebox{0}{\resizebox{.45\textwidth}{!}{\includegraphics{scalingt2.eps}}}
    \caption{(Color online). $G(\vec r,t,t)$ is plotted against $x=r/L(t)$ for a quench to 
             $T=0$ (left) or $T=2$ (right). Different curves correspond to 
             several times (the same in the two figures) 
             between $t=10$ and $t= 10^4$ (from bottom to top,
             for $T=2$). The dot-dashed line
             is the Porod law $y=1-ax$, where $a$ (the same in the two
             pictures) is obtained as the best fit of $G(\vec r,t,t)$ at $T=0$.
             Points are joined by a piecewise continuous line as a guide
             for the eye (for clarity the first two points, the one in the origin and
             the following, are not joined).}
\label{figporod}
\vspace{2cm}
\end{figure}

The interplay between $U(t,T_f)$ and $L(t)$ can also be observed in the
behavior of the equal time correlation function, which, when scaling
holds, should behave as in Eq.~(\ref{acca}),
with $h$ obeying Porod law~(\ref{porod})
in the case of sharp interfaces. However, 
as already discussed, the presence of $U(t,T_f)$ introduces a correction 
to scaling in an early regime 
when $L(t)$ has not yet grown sufficiently larger than $U(t,T_f)$.
Moreover, due to roughness, interfaces are not sharp. Then,
both scaling  and the Porod law are expected to be violated 
for $r\lesssim U(t,T_f)$, that is for 
$x=r/L(t)\lesssim x_R(t)=U(t,T_f)/L(t)=a_2(T_f)L^{-1/2}(t)$,
namely in a range of $x$ that shrinks in time but that may  be 
appreciable for large $T_f$. Actually this is observed in Fig.~\ref{figporod}.
While curves of $G_{ag}(r,t,t)$ for different times collapse when plotted against $r/L(t)$
for the quench to $T_f=0$, as discussed in Sec.~\ref{numteq0},
and the Porod law is also verified, 
when the quench to $T_f>0$ is considered one 
observes significant scaling violations in the region of small $x$.
In this regime, the curve definitely deviates from the linear Porod law.
As time goes on, these violations become weaker and the curves seem to
approach the same behavior as for $T_f=0$. For intermediate temperatures similar,
but less pronounced, violations are also observed.

\section{Summary and conclusions} \label{concl}

In this paper we have investigated the role of $T_f$ in the phase-ordering kinetics of the
Ising model with single spin flip dynamics. At $T_f=0$ the dynamics 
is characterized by faceted interfaces and by the kinetic constraint
of the impossibility of activated moves.
At $T_f>0$ interfaces are curved and rough (for $T>T_R$ in $d=3$). 
We have shown that, while the exponent $z$ regulating the decay of the total 
density of interfacial spins is not changed by the different geometry of
interfaces in quenches to $T_f=0$ or $T_f>0$, 
other quantities, such as the exponents $\beta _n$ describing the behavior of
the densities of particular classes of spins do change. 
The existence of two different dynamical mechanisms induces a
crossover pattern for finite $T_f$.    
Besides, in $d=2$ for $T_f>0$ the roughening length competes with
$L(t)$ in an early stage, delaying the realization of dynamical scaling,
as it is evidenced by the time dependence of the effective exponent
$z_{eff}$ and by the breakdown of the Porod law at small $r/L(t)$.

This whole pattern of behaviors is due to equilibrium properties of 
the interfaces. Therefore we expect to observe similar behaviors
for the Ising model with conserved dynamics.

\vspace{1cm}
{\large {\bf APPENDIX I}}
\vspace{1cm}

We consider a generic profile of the $h_i $ and denote with $\{h \}$ this configuration. 
Let us introduce the probability $P(\{h \},t)$ of having such a configuration
at time t, and the conditional probability $P(\{h'   \},t'\vert \{h \},t)$
of having $\{h'\}$ at time $t'$ provided that the configuration $\{h \}$
was found at $t<t'$. One has
\be
\langle h_i (t+\Delta t)-h_i(t)\rangle =\sum _{\{h\},\{h'\}}(h'_i-h_i)
P(\{h'\},t+\Delta t \vert \{h\},t)P(\{h \},t).
\label{difffin}
\ee
From the master equation, the conditional probability can be written as~\cite{Lippiello05}
\be
P(\{h'\},t+\Delta t \vert \{h \},t)=\left [ 1-\Delta t \sum _j 
\sum _{h''_j\neq h_j}w(h_j\to h_j'')\right ]\delta _{\{h'\},\{h\}}+
\Delta t \sum _j \sum _{h''_j\neq h'_j}w(h_j''\to h_j')
\delta _{\{h''\},\{h\}}+{\cal O}(\Delta t ^2)
\label{probcond}
\ee
where we have introduced the transition rate $w(h_j\to h_j')$ 
for moving the height $h_j$ of the $j$-th column to $h_j'$
and $\delta _{\{h''\},\{h\}}$ is the Kronecker function between
configurations $\{h''\}$ and $\{h\}$.
We have assumed that the $w$ 
are single spin flip transition rates
and hence transitions between configurations can be obtained by summing over $j$.
Apart from this, the $w$ are still generic (not necessary those of the
zero temperature Ising model) at this stage.
Inserting Eq.~(\ref{probcond}) into Eq.~(\ref{difffin}), due to the term
$(h'_i-h_i)$ only contributions with $\{h'\}\neq \{h\}$ and $j=i$ survive, and one has
\be
\langle h_i (t+\Delta t)-h_i(t)\rangle =
  \Delta t \sum _{\{h \},\{h'\}\neq \{h\}}
  (h'_i-h_i)P(\{ h \},t)w(h_i\to h_i').
\ee
Since in an elementary move $h_i\to h_i\pm 1$ is only allowed, introducing $m=\pm 1$
the last Equation reads
\be
\langle h_i (t+\Delta t)-h_i(t)\rangle =\Delta t \sum _{\{h \}}\sum _m
m w(h_i\to h_i+m)P(\{ h \},t).
\ee
Taking the continuum limit $\Delta t \to 0$ yields
\be
\frac{d\langle h_i (t)\rangle}{dt}=\sum _{\{h \}}\sum _m
m w(h_i\to h_i+m)P(\{ h \},t).
\label{contlim}
\ee
This equation has been obtained without any approximation. We now have to specify
the form of the $w(h_i\to h_i+m)$ in order to reproduces the original rules of 
the zero temperature Ising model. However, in order to have a 
tractable model, we consider transition rates
which correspond to the Ising model with 
the additional constraint 
\be 
\vert h_{i+1}(t)-h_i(t)\vert \leq 1 \quad \forall i, 
\label{constraint} 
\ee
as discussed in Sec.~\ref{relaxtime}.
Starting with
the case $d=2$, we define them as
\be
w(h_i\to h_i+m)=\frac{m}{2}\nabla ^2h_i+F_i^{(m)}(\{ h\}),
\label{w}
\ee
where $\nabla ^2h_i = h_{i-1}+h_{i+1}-2h_i$ is the discrete Laplacian 
in one dimension, and
\be
F_i^{(m)}(\{ h\})=\frac{1}{2}\vert \nabla ^2 h_i\vert 
\delta _{m \, sign \left (\nabla ^2 h_i\right ),\,-1}.
\label{ww}
\ee
In order to see this let us notice first that if the configuration $\{h\}$
satisfies the condition~(\ref{constraint}),
with the transition rates~(\ref{w}) that constraint will never be violated
by the later evolution. In fact, let us focus on site $i$ and suppose that
the transition $h_i\to h_i+1$ is going to be attempted. This transition
would violate Eq.~(\ref{constraint}) if $h_{i+1}=h_i-1$. In this case, however,
it easy to check that for every $h_{i-1}$ consistent with the condition~(\ref{constraint})  
it is $w(h_i\to h_i+1)=0$. The same argument can be repeated for every configuration.
Having proved that ~(\ref{constraint}) is fulfilled by the transition 
rates~(\ref{w})
we can restrict ourselves to consider the only cases allowed, namely those with 
$\nabla ^2 h_i=0,\pm 1,\pm 2$. When $\nabla ^2 h_i=0$ the only configuration
where the move $h_i\to h_i\pm 1$ could be attempted without violating the
constraint~(\ref{constraint}) is that with $h_{i-1}=h_i=h_{i+1}$. In this case,
in the original Ising model the move is forbidden, which agrees with
Eq.~(\ref{w}) giving $w(h_i\to h_i+m)=0$ in this case. Coming to the cases
with $\nabla ^2 h_i=\pm 1$ (which are realized, for instance, when
$h_{i+1}=h_i=h_{i-1}\mp 1$),
in the original Ising model moves with $m\,\,sign\left (\nabla ^2 h_i\right )<0$
are energetically forbidden, while those with $m\,\,sign \left (\nabla ^2 h_i\right )>0$
occur with a probability $1/2$, using Glauber transition rates. This agrees
with Eq.~(\ref{w}). Analogously, when $\nabla ^2 h_i=\pm 2$ (when, for instance,
$h_{i+1}=h_{i-1}$ and $h_i=h_{i+1}\mp 1$),
in the original Ising model moves with $m\,\,sign\left (\nabla ^2 h_i\right )<0$
are forbidden, while those with  $m\,\,sign\left (\nabla ^2 h_i\right )>0$
lower the energy and occur with a probability $1$, providing again agreement
with Eq.~(\ref{w}). 

Let us now insert the transition rates~(\ref{w}) into the evolution 
Equation~(\ref{contlim}), obtaining
\be
\frac{d\langle h_i (t)\rangle}{dt}=\sum _{\{h \}}\sum _m
\frac{m^2}{2} \nabla ^2 h_i P(\{ h \},t)+
\sum _{\{h \}}\sum _m mF_i^{(m)}(\{ h \})P(\{ h \},t).
\label{quasi}
\ee
Performing the sum over $m$ one has
\be
\frac{d\langle h_i (t)\rangle}{dt}=\langle \nabla ^2  h_i \rangle
+\langle F^\pm_i \rangle
\label{quasiquasi}
\ee
where $F_i^\pm(\{ h\})=F_i^{(1)}(\{ h\})-F_i^{(-1)}(\{ h\})$.
Let us now consider the last term on the r.h.s. of Eq~(\ref{quasiquasi}).
We want to show that it can be neglected.
We will show it separately, for all the possible values of $\nabla ^2 h_i$, namely $\nabla ^2 h_i=0,\pm1,\pm2$.
For $\nabla ^2 h_i=0 $, from Eq.~(\ref{ww}), it is $F_i^\pm(\{h\})\equiv 0$. 
Let us consider now the contributions  
with $\nabla ^2 h_i=\pm 1$. This situation corresponds
to spins with $n=0$, or steps in the terminology of Sec.~\ref{densities}.
These that can be flipped from $\sigma =1$ to 
$\sigma = -1$ and back without energy costs. Therefore the two values
of the spin in this case occur with equal probability (i.e. 1/2) and,
as $\nabla ^2 h_i$
changes its sign when the spin is reversed the contributions $\nabla ^2 h_i=\pm 1$ cancel in
$\langle F^\pm_i \rangle$. The case $\nabla ^2 h_i=- 2$ can never be realized in the kinetics 
because it would require a move with energy increase.
Therefore, the terms with $\nabla ^2 h_i=0,\pm1,-2$ do not contribute
to the r.h.s. of Eq.~(\ref{quasiquasi}). 
This is no longer true for $\nabla ^2 h_i=2$. This term
corresponds, in the language of Sec.~\ref{densities}, 
to a spin with $n=-2$ which, as explained 
in Sec.~\ref{densities}, are created only when 
the last spin has to be reversed to complete a row. 
On average
this happens once every $l$ moves. Therefore, although the
contributions with $\nabla ^2 h_i=2$ do not strictly vanish, 
they provide a contribution $\langle F^\pm_i \rangle \propto 1/l$,  which can be neglected 
in the large-$l$ limit. Then we arrive at the diffusion equation
\be
\frac{d\langle h_i (t)\rangle}{dt}=\langle \nabla ^2 h_i \rangle
\label{diffusion}
\ee 

Let us turn now to the case $d=3$. 
Similarly to the case $d=2$, 
it is easy to check that the transition rates
\be
w(h_i\to h_i+m)=\frac{m}{4}\nabla ^2h_i+F_i^{(m)}(\{ h\}),
\label{w3}
\ee
with the following form of $F_i^{(m)}(\{h \})$
\be
F_i^{(m)}(\{ h\})=\frac{1}{4}\vert \nabla ^2 h_i\vert 
\delta _{m \, sign \left (\nabla ^2 h_i\right ),\,-1} +
f _i(\{ h\})\delta _{m \, sign \left (\nabla ^2 h_i\right ),\,1}, 
\ee
where $\nabla ^2 h_i$ is the discretized laplacian in $d=3$ and
with the values of $f_i (\{h\})$ given in table~\ref{tab},
satisfy the constraint~(\ref{constraint}) and reproduce the Glauber transition
rates of the original Ising model at $T=0$.
Proceeding as for $d=2$ one arrives at 
\be
\frac{d\langle  h_i (t)\rangle}{dt}=\frac{1}{2}\langle \nabla ^2 h_i \rangle
+\langle F^\pm _i\rangle,
\ee
By reasoning as in $d=2$, one concludes that only sites with 
$\vert \nabla ^2 h_i \vert= 1,2,3,4$, corresponding to spins with $E_i\neq 0$
contribute to $\langle F^\pm _i\rangle$, but these can be neglected
for large $l$. 
Hence one arrives also in this case to a diffusion equation
\be
\frac{d\langle h_i (t)\rangle}{dt}=\frac{1}{2}\langle \nabla ^2 h_i \rangle.
\ee

\vspace{2cm}
\begin{table}
\begin{center}
\begin{tabular}{|c|c|}
\hline
$\mid \nabla ^2 h_i\mid$ & $f_i(\{h\})$ \\
\hline
\hline
0 & 0\\
\hline
1 & -1/4 \\
\hline
2 & $\left \{ \begin{array}{ll}
	-1/2 \quad for \quad \nabla ^2 _x=0 \,\, or \,\, \nabla ^2 _y=0 \\
	0 \quad \quad \quad else \\
        \end{array} 
        \right .$ \\
\hline
3 & 1/4 \\
\hline
4 & 0 \\
\hline
\end{tabular}
\end{center}
\caption{The values of the function $f_i(\{h\})$.  
$\nabla ^2_x$ ($\nabla ^2_y$) is the discrete second derivative along
$x$ ($y$).}
\label{tab}
\end{table}

\vspace{1cm}
{\large {\bf APPENDIX II}}
\vspace{1cm}

\begin{figure}
\vspace{2cm}
    \centering
   \rotatebox{0}{\resizebox{.5\textwidth}{!}{\includegraphics{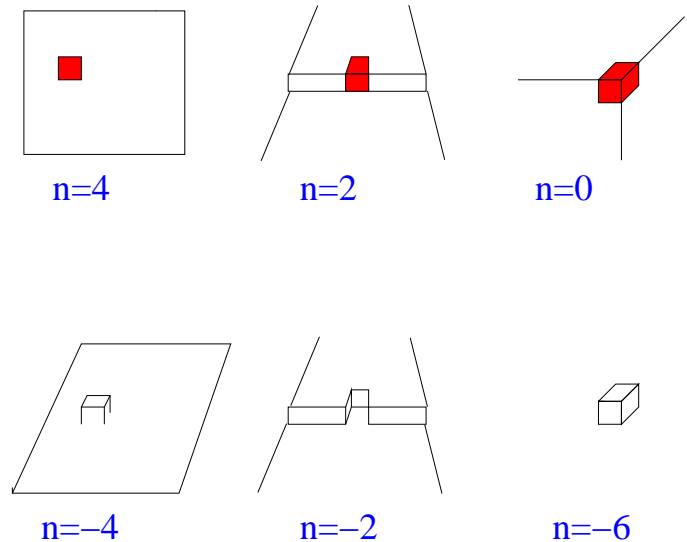}}}
    \caption{(Color online). Classification of interfacial spins.}
\label{class_spins3d}
\vspace{2cm}
\end{figure}

Let us consider the shrinkage of a cubic bubble of linear size $l$,
and extend the argument developed in Sec.~\ref{densities} to the case $d=3$. 
The interfacial spins can be classified according to $n$,
as shown in Fig.~\ref{class_spins3d}. 
Suppose again that the interface grows from the bottom of the bubble.
Let us define $h_i$ as the height of the $i-$th column,
with $i=1,...,l^2$  running on the two-dimensional lattice.
While columns are growing, the profile of $h_i$ is made of
flat parts (spins with $n=4$), edges (spins with n=2) and corners 
(spins with $n=0$). We make again the hypothesis that the probability
of their occurrence is finite and constant. By reasoning analogously
to the $d=2$ case, since these spins belong 
to the growing surface,
their number is proportional to $l^2$. Hence
$\rho _4(l)\propto \rho _2(l)\propto \rho_0 (l)\propto l^{-1}$.
Let us imagine
to reverse the spins from the bottom, level by level. 
A spin with $n=-4$ is only produced when the last spin of a certain
level has to be reversed.
All the $l$ levels are completed in a time $\tau _l\propto l^2$,
so that there is a number proportional to $l^{-1}$ of spins 
with $n=-4$ in a unit time.  This implies $\rho _{-4}(l)\propto l^{-4}$.
Spins with $n=-2$ are generated analogously to the spins with
$n=-2$ in $d=2$, namely when the last spin must be reversed
in order to complete a row. 
$l^2$ rows must be completed in a time $\tau \propto l^2$ in order
to reverse all the spins of the bubble.
Therefore there are $l^{0}$ spins with $n=-2$
in a unit time. In conclusion $\rho _{-2}(l)\propto l^{-3}$.
Finally, spins with $n=-6$ are only formed when
a growing column reaches the top of the bubble, with the condition
that all the nearest column have already reached the top. 
Due to the kinetic constraints, these neighboring
column must themselves have at least one nearest column 
of equal or higher eight. One can iterate 
this argument until a border is reached. Since the border is a
distance of order $l$ away, one concludes that
every $l$ spins only one can have $n=-6$. Therefore
a number proportional to $l$ of spins are generated when
the columns reach the top, and this event happens once
every $\tau _l\propto l^2$ moves. Then $\rho _{-6}(l)\propto l^{-4}$.
Assuming scaling, and identifying $l$ with $L(t)$ in the 
phase-ordering kinetics one arrives at Eqs.~(\ref{alfaf3d}-\ref{alfai3d}).


\begin{thebibliography}{99}

\bibitem{Bray94}
For a review see A.J.~Bray, Adv.Phys. {\bf 43}, 357 (1994).

\bibitem{Furukawa}
H.~Furukawa, J.Stat.Soc.Jpn. {\bf 58}, 216 (1989); Phys.Rev. B {\bf 40}, 2341 (1989).

\bibitem{Fisher88}
D.S.~Fisher and D.A.~Huse, Phys. Rev. B {\bf 38}, 373 (1988);

\bibitem{Mason93}
N.~Mason, A.N.~Pargellis and B.~Yurke, Phys. Rev. Lett. {\bf 70}, 190 (1993);
F.~Liu and G.F.~Mazenko, Phys. Rev. B {\bf 44}, 9185 (1991).

\bibitem{hohenberg}
P.C.~Hohenberg and B.I.~Halperin, Rev. Mod. Phys. {\bf 49}, 435 (1977).

\bibitem{mazenkorg}
G.F.~Mazenko and O.T.~Valls Phys. Rev. B {\bf 26}, 389 (1982);
G.F.~Mazenko, O.T.~Valls, and F.C.~Zhang, Phys. Rev. B {\bf 31}, 4453 (1985);
Phys. Rev. B {\bf 32}, 5807 (1985);	
Z.W.~Lai, G.F.~Mazenko, and O.T.~Valls, Phys. Rev. B {\bf 37}, 9481 (1988);
C.~Roland and M.~Grant, Phys. Rev. B {\bf 39}, 11971 (1989);
Phys. Rev. Lett. {\bf 60}, 2657 (1988).

\bibitem{Bray89}
A.J.~Bray, Phys. Rev. Lett. {\bf 62}, 2841 (1989); Phys. Rev. B {\bf 41}, 6724 (1990).

\bibitem{sicilia}
Limited to the case $d=2$, some exact result have been derived
in: 
J.J.~Arenzon, A.J.~Bray, L.F.~Cugliandolo and A.~Sicilia, Phys. Rev. Lett. {\bf 98}, 
145701 (2007); A.~Sicilia, J.J.~Arenzon, A.J.~Bray, L.F.~Cugliandolo,
Phys. Rev. E {\bf 76}, 061116 (2007). 

\bibitem{Allen79}
S.M.~Allen and J.W.~Cahn, Acta Metall. {\bf 27}, 1085 (1979);
I.M.~Lifshitz, Zh. Eksp. Teor. Fiz. {\bf 42}, 1354 (1962) [Sov. Phys. - JETP 
{\bf 15}, 939 (1962).

\bibitem{Brown02}
G.~Brown and P.A.~Rikvold, Phys. Rev. E {\bf 65}, 036137 (2002).

\bibitem{Amar89}
J.G.~Amar and F.~Family, Bull. Am. Phys. Soc. {\bf 34}, 491 (1989);
J.D.~Shore, M.~Holzer and J.P.~Sethna, Phys. Rev. B {\bf 46}, 11376 (1992).

\bibitem{nota1}
The scaling approach we develop here recalls that of
Ref.~\cite{Bray98}, where the exponent $z$ is determined
by evaluating the relaxation time of a distorted interface,
but in the framework of continuum theories.

\bibitem{Cahn}
J.W.~Cahn, Acta Metall. {\bf 9}, 795 (1961); Acta Metall. {\bf 10}, 179 (1962);
Acta Metall. {\bf 14}, 1685 (1966); Trans. Metall. Sec. AIME {\bf 242}, 166.

\bibitem{nota4}
By equilibrium interface we mean the stationary state of an interface
in an Ising system with antiperiodic boundary condition
in one direction. 

\bibitem{Corberibis}
R.~Burioni, D.~Cassi, F.~Corberi, and A. Vezzani, Phys. Rev. E {\bf 75}, 011113 (2007);
F.~Corberi, E.~Lippiello, and M.~Zannetti, Phys. Rev. E {\bf 74}, 041106 (2006);
R.~Burioni, D.~Cassi, F.~Corberi, and A.~Vezzani, Phys. Rev. Lett. {\bf 96}, 235701 (2006);
F.~Corberi, E.~Lippiello and M.~Zannetti, Phys.Rev. E {\bf 68}, 046131 (2003); 
F.~Corberi, E.~Lippiello, and M.~Zannetti, Phys. Rev. E {\bf 63}, 061506 (2001);
F.~Corberi, E.~Lippiello, M.~Zannetti, Eur. Phys. J. B {\bf 24} (2001), 359.

\bibitem{Corberi03}
F. Corberi, E. Lippiello and M. Zannetti, Phys. Rev. E 72, 056103 (2005).


\bibitem{Manoy00}
G.~Manoy and P.~Ray, Phys. Rev. E {\bf 62}, 7755 (2000);
E.~Lippiello, F.~Corberi, and M.~Zannetti, Phys.Rev.E {\bf 74},
041113 (2006). 


\bibitem{Barabasi}
A.-L. Barab\'asi and H. E. Stanley, {\it Fractal Concepts in Surface Growth},
(Cambridge University Press, Cambridge, 1995).

\bibitem{Abraham89}
The argument is quoted in D. B. Abraham and P. J. Upton, 
Phys. Rev. B {\bf 39}, 736 (1989).

\bibitem{Lippiello05}
E.~Lippiello, F.~Corberi, and M.~Zannetti, Phys.Rev.E {\bf 71},
036104 (2005). 

\bibitem{Bray98}
A.J.~Bray, Phys. Rev. E {\bf 58}, 1508 (1998).


\end{thebibliography}
\end{document}